\documentclass[useAMS,usenatbib]{mnras}
\usepackage{color}
\usepackage{graphicx}
\usepackage{lscape}
\usepackage{amsmath}

\usepackage{float}

\begin{document}
\title[Spectroscopic + photometric analysis of galaxies with {\sc starlight}]{
Simultaneous spectroscopic and photometric analysis of galaxies with {\sc starlight}:
CALIFA $+$ GALEX
}

\author[L\'opez Fern\'andez et al]
         {L\'opez Fern\'andez, R.$^{1}$\thanks{E-mail: 
rafael@iaa.es},
      Cid Fernandes, R.$^{2}$,
	  Gonz\'alez Delgado, R. M.$^{1}$,
	  \newauthor
	  Vale Asari, N.$^{2}$,
	  P\'erez, E.$^{1}$,
 	  Garc\'ia-Benito, R.$^{1}$,  
	  de Amorim, A. L.$^{2}$,
	  \newauthor
	  Lacerda, E. A. D.$^{2}$,
	  Cortijo-Ferrero, C.$^{1}$,
	  and S\'anchez, S. F.$^{3}$\\
	 $^{1}$Instituto de Astrof\'{\i}sica de Andaluc\'{\i}ıa (CSIC), P.O. Box 3004, 18080 Granada, Spain\\
	 $^{2}$Departamento de F\'{\i}sica - CFM - Universidade Federal de Santa Catarina,
	 Florian\'opolis, SC, Brazil\\
	 $^{3}$Instituto de Astronom\'ia,Universidad Nacional Auton\'oma de M\'exico,
A.P. 70-264, 04510 M\'exico D.F., M\'exico
	 }

\maketitle

%\pagerange{\pageref{firstpage}--\pageref{lastpage}} \pubyear{2009}

\begin{abstract}
We present an extended version of the spectral synthesis code {\sc starlight} designed to incorporate  both $\lambda$-by-$\lambda$ spectra  and photometric fluxes in the estimation of stellar population properties of galaxies. 
The code is tested with simulations and data for 260 galaxies culled from the CALIFA survey,  spatially matching the 3700--7000 \AA\ optical datacubes to GALEX  near and far UV images. 
The sample spans E--Sd galaxies with masses from $10^9$ to $10^{12} M_\odot$ and stellar populations all the way from star-forming to old, passive systems. Comparing results derived from purely optical fits with those which also consider the NUV and FUV data we find that: 
(1) The new code is capable of matching the input UV data within the errors while keeping the quality of the optical fit essentially unchanged. 
(2) Despite being unreliable predictors of the UV fluxes, purely optical fits yield stellar population properties which agree well with those obtained in optical+UV fits for nearly 90\% of our sample.
(3) The addition of UV constraints has little impact on properties such as stellar mass and dust optical depth. Mean stellar ages and metallicities also remain nearly the same for most galaxies, the exception being low-mass, late-type galaxies, which become older and less enriched due to rearrangements of their youngest populations. 
(4) The revised ages are better correlated with observables such as the 4000 \AA\ break index, and the $NUV - r$ and $u - r$ colours, an empirical indication that the addition of UV constraints helps mitigating the effects of age-metallicity-extinction degeneracies.
\end{abstract}

\begin{keywords}
techniques: spectroscopic -- techniques: photometric -- galaxies: evolution
\end{keywords}

\label{firstpage}

\section{Introduction}
\label{sec:Intro}

Converting photometric or spectroscopic data on galaxies into physical quantities is part of the daily routine in extragalactic work. It is this translation that produces  estimates of the mass in stars, their typical ages and metallicities,  
dust attenuation and other properties which ultimately mould our current understanding of galaxies and their stellar content. 

This process works upon data that often consist of multi-band photometry, preferably covering as much of the spectral energy distribution as possible, say GALEX plus SDSS (e.g. Kaviraj et al. 2007a; Salim et al. 2007; Schiminovich et 
al. 2007; Schawinski et al. 2014) or 2MASS magnitudes (e.g. Barway et al. 2013). Other times the input is spectroscopic, such as optical spectra provided by the SDSS. Methods to analyse $F_\lambda$  split into those which reduce $F_\lambda$ to a set of spectral indices (say,  Lick indices) and those which attempt to fit it $\lambda$-by-$\lambda$ -- see Walcher et al.\ (2011) for a comprehensive review. Because of the informative nature of absorption and emission lines spectroscopy is in principle more powerful than photometry. Purely spectroscopic studies,  however,  are invariably limited in $\lambda$-coverage, hence limited in the ability to exploit  stellar population information encoded over long $\lambda$ baselines more easily covered with multi-band photometry.
To quote a known example  which will appear later in this paper, low level ongoing star formation may leave weak/undetectable imprints in the optical continuum of a galaxy while at the same time accounting for most of its UV flux. Clearly, a combined optical $+$ UV analysis would lead to better estimates of a galaxy's star formation history (SFH).
This is precisely the goal of this paper. 

While tools to retrieve stellar population properties out of multi-band photometry are common in the literature (e.g. the CIGALE code of Noll et al.\  2009), methods which mix photometric and spectroscopic information are less common. An example is the work of Schawinski et al.\ (2007), who model UV to IR photometry in conjunction with Lick indices derived from SDSS spectra. Our goal in this paper is to go a step further and, instead of summarizing the spectroscopic information by a set of  indices, combine photometry with a full $\lambda$-by-$\lambda$ analysis.

We develop and test a combined spectroscopic $+$ photometric analysis built upon the full spectral fitting code {\sc starlight} of Cid Fernandes et al.\ (2005). The method admits any combination of spectra and photometry, but we focus on the specific case of CALIFA optical spectroscopy (S\'anchez et al.\ 2012) plus GALEX photometry (Martin et al.\ 2005), a combination which has the key advantage of allowing us to mitigate aperture effects, a serious source of concern in any experiment involving data gathered through different instruments. This `upgrade' is part of a more ambitious program to include panchromatic constraints in {\sc starlight}, like far-IR data (to constrain the dust reprocessed luminosity), and recombination emission line fluxes (tracers of the $h\nu > 13.6$ eV field) or ratios (sensitive to dust attenuation). 

We organize this article as follows. Section \ref{sec:Method} describes how to combine spectra and photometry into a single figure of merit to be optimized. This entails both `philosophical' issues, like how much weight one wishes to attribute to each kind of data, as well as more technical aspects, such as whether the photometry needs to be treated as upper or lower limits due to, say, aperture mismatches with respect to the spectroscopic data. Section \ref{sec:Simulations} presents simulations designed to test the code under realistic circumstances. Section \ref{sec:sample} explores an actual application based on a combination of optical spectra from  CALIFA data and GALEX fluxes. Section \ref{sec:Discussion} discusses how the addition of UV constraints improves the estimation of ages and metallicities, and the implications for the stellar mass-metallicity relation. Our main results are summarized in  Section \ref{sec:Summary}.

\section{Method}
\label{sec:Method}

This section presents the basic formalism used throughout the paper. Our combined spectroscopic $+$ photometric modelling can in principle be applied to any arbitrary combination of data, like a near-IR spectrum plus optical and/or UV photometry, an optical spectrum plus near-IR photometry, etc. For concreteness, the simulations and actual applications explored in later sections focus on the combined analysis of a 3700--7000 \AA\ optical spectrum and NUV ($\sim 2274$ \AA) plus FUV ($\sim 1542$ \AA)  photometry from GALEX. 

%The data can pertain either to a galaxy as a whole or to some region of it. The spatial correspondence between the spectroscopic and photometric data, as well as the relative weight given to the two datasets have to be handled with care, and we discuss ways to handle these issues.

\subsection{Input data}

We describe the spectroscopic data to be fitted in terms of the following elements:

\begin{enumerate}

\item The observed spectrum $O_\lambda$ and its error $\sigma(O_\lambda)$.

\item Mask (mask$_\lambda$) and flag (flag$_\lambda$) spectra to mark regions to be discarded from the analysis because of emission lines or artifacts (e.g. bad pixels and sky residuals). 

\item $w_\lambda = \sigma(O_\lambda)^{-1}$ is the weight given to pixel $\lambda$. Masked and flagged pixels have $w_\lambda = 0$. Discounting zero-weight entries one is left with  $N^{\rm eff}_\lambda$ fluxes to be fitted.  $N^{\rm eff}_\lambda$ is typically of the order of $10^3$.

\end{enumerate}

The photometric data, indexed with a subscript $l$ running from $l = 1$ to $N_l$ filters, consist of

\begin{enumerate}

\item The apparent AB magnitude of the object $m^{\rm obs}_l$ and its error $\sigma(m^{\rm obs}_l)$. 

\item The filter transmission curves $T_l(\lambda)$.

\end{enumerate}

The input photometry is corrected for Galactic extinction, but K-corrections are not necessary since we will perform the synthetic photometry in the galaxy's redshift $z$.

\subsection{The model}

Model predictions for the observed spectroscopic and photometric fluxes are built by a linear combination of spectra from a base $B_{j,\lambda}$ ($j = 1 \ldots N_\star$), usually (but not necessarily) drawn from evolutionary synthesis models for simple stellar populations (SSP) of different ages and metallicities. Each base spectrum is first normalized by its value at a chosen reference wavelength $\lambda_0$ ($ = 5635$ \AA\ in this paper). The scaled base spectra are then combined in proportions $x_j$ to build a model spectrum $M_\lambda$ given by

\begin{equation}
\label{eq:ModelFlux}
M_{\lambda} = 
M_{\lambda_{0}}  \left( \sum_{j=1}^{N_{\star}} x_{j} b_{j,\lambda} \right)
r_\lambda
\otimes G(v_{\star},\sigma_{\star})
\end{equation}

\noindent where $b_{j,\lambda} \equiv B_{j,\lambda} / B_{j,\lambda_0}$, $r_\lambda$  is a shorthand for the $e^{-\tau_V (q_\lambda - q_{\lambda_0})}$ reddening produced by a foreground screen of dust with an extinction curve $q_\lambda = \tau_\lambda / \tau_V$, and $G(v_\star,\sigma_\star)$ denotes a gaussian kinematical kernel centred at velocity $v_\star$ and with dispersion $\sigma_\star$.

Provided the base  spectra cover the wavelengths of our $N_l$ filters, 
Eq.\ (\ref{eq:ModelFlux}) can be used to predict model magnitudes $m_l$

\begin{equation}
\label{eq:ModelMag}
%m_l(\vec{x},\tau_V,v_\star,\sigma_\star, z) =
m_l = -2.5\log\frac{\displaystyle\int M_{\lambda / (1 + z)} T_l(\lambda)\lambda\, d\lambda}{\displaystyle\int T_l(\lambda)\lambda^{-1}\, d\lambda} - 2.41
\end{equation}

\noindent where the rest-frame model spectrum $M_\lambda$ is shifted to the observed frame, thus  circumventing the need for K-corrections.

\subsection{Combining spectroscopic and photometric figures of merit}

A purely spectroscopic analysis would consist of, for instance, estimating the model parameters by minimizing  

\begin{equation}
\label{eq:chi2_SPEC}
\chi^2_{\rm SPEC} = \sum_\lambda w_\lambda^2 (O_\lambda - M_{\lambda})^2
\end{equation}

\noindent as in Cid Fernandes et al.\ 2005. The analogous photometric figure of merit is

\begin{equation}
\label{eq:chi2_PHO}
\chi^2_{\rm PHO} = \sum_{l=1}^{N_l} \left(  \frac{ m^{\rm obs}_l - m_l }{ \sigma_l } \right)^2
\end{equation}

\noindent which compares model and observed magnitudes. The total $\chi^2$ to be considered in a joint analysis is then simply 

\begin{equation}
\label{eq:chi2_TOT}
\chi^2_{\rm TOT} = \chi^2_{\rm SPEC} + \chi^2_{\rm PHO} 
\end{equation}

We have also experimented with other definitions of $\chi^2_{\rm TOT}$. One might, for instance, want to ensure that the spectroscopic and photometric data be given commensurable weights in the joint analysis. This can be implemented by scaling $\chi^2_{\rm PHO}$ by a factor of $\sim N_{\lambda}^{\rm eff} / N_l$ (or, equivalently, scaling the observational errors). 

Throughout this work we adopt Eq.\ \ref{eq:chi2_TOT} in its original form. The simulations and real-data applications presented below produced good spectroscopic and photometric fits with no need for {\it ad hoc} scaling factors. Such weighting schemes should be more relevant in cases where the observational errors are not well known.

\subsection{Aperture mismatch: Implementation of photometric constraints as ranges}

Another problem to consider when modelling spectra and photometry is that they are often collected through different apertures. The actual applications explored in Section \ref{sec:sample} are based on a combination of optical spectra from  CALIFA data and GALEX fluxes, whereby the spectra are extracted from the projection of the GALEX aperture on the integral field data cube, thus mitigating aperture effects. Nonetheless, in the interest of completeness and future reference we discuss the `range-fitting' scheme implemented in the new {\sc starlight} to deal with aperture uncertainties.

Even disregarding spatial variations of the stellar populations, an aperture mismatch implies flux scale differences which make the simultaneous fit of $O_\lambda$ and $m_l$ meaningless. Aperture corrections are designed to fix this problem by scaling the input data to a same flux level, yet these are but approximate corrections, subject to systematic uncertainties. 

To deal with this issue we introduce a modified version of $\chi^2_{\rm PHO} = \sum_l \chi^2_l$ where the $e^{-\chi^2_l/2}$ gaussian likelihood of each $m^{\rm obs}_l$  implicit in Eq.\ (\ref{eq:chi2_PHO}) is replaced by a flat top gaussian likelihood, where $\chi^2_l$ is given by 

\begin{equation}
\chi^2_l = \left\{
  \begin{array}{cr}
    \left(\frac{m_l - m_l^{\rm low}}{\sigma_l} \right)^2  &  m_l \le m_l^{\rm low}\\
    0                                              &  m_l^{\rm low} < m_l < m_l^{\rm upp}\\
    \left(\frac{m_l - m_l^{\rm upp}}{\sigma_l} \right)^2  &  m_l \ge m_l^{\rm upp}\\
  \end{array}
\right.
\end{equation}

This modification acts in the sense that the model magnitude $m_l$ no longer sees $m^{\rm obs}_l$ as a 
target to be matched within a $\sim \pm \sigma_l$ margin of error. Instead, the code will seek 
$m_l$ values which do not depart by much more than $\sim \sigma_l$ from the $m_l^{\rm low} < m_l < m_l^{\rm upp}$ range. For $\sigma_l \ll (m_l^{\rm upp} -  m_l^{\rm low})$ the likelihood effectively becomes a box car, which guarantees that the model $m_l$ remains within the allowed range.  Put another way, all solutions leading to $m_l$ in the given $m_l^{\rm low}$--$m_l^{\rm upp}$ range are equally acceptable, contributing nothing to the global figure of merit. 

A qualitatively similar effect could be obtained by exaggerating the uncertainty in $m_l$ beyond its nominal value $\sigma(m_l)$. We nevertheless prefer the `range-fitting' recipe outlined above, which has the advantage of expressing in an explicit way the systematic character of uncertainties in matching fluxes from different instruments/telescopes. An added benefit of our formulation is that it allows the incorporation of lower or upper limits in the analysis. Suppose that all we know about the FUV flux in a galaxy is that it is weaker than $m_{\rm FUV}^{\rm upp}$. One can couple this observational upper limit to an arbitrarily low lower limit and feed this information into our recipe to enforce that the resulting stellar populations will conform to the given upper limit. As shown in Section \ref{sec:Simulations}, purely optical studies can easily allow optically insignificant but UV-dominant populations, so that the use of UV limits can be useful.

As mentioned above, our data allow us to match the $O_\lambda$ and $m_l$ apertures, so that no range-fitting scheme is necessary. We thus set $m_l^{\rm low} = m_l^{\rm upp} = m^{\rm obs}_l$, so that the code tries to fit the observed magnitude within its observational error.

\subsection{Spectral base}
\label{sec:Base}

The spectral base $B_{j,\lambda}$ is the key astrophysical ingredient in our analysis, providing the translation of photometric and spectroscopic observables into stellar population  properties. In this study we work with a base comprised by $N_\star = 246$ SSPs with ages from $t = 1$ Myr to 14 Gyr and metallicities in the $Z = 0.005$--$2.5 Z_\odot$ range drawn from a preliminary version of  an update to the Bruzual \& Charlot (2003) models (G. Bruzual 2007, private communication). 
This same set of models was called base {\it CBe} by Gonz\'alez Delgado et al.\ (2015), which describes it more thoroughly. The initial mass function is that of Chabrier (2003).

These models adequately cover all spectral ranges addressed in this study, namely optical and UV. In fact,
the requirement of optical--UV coverage eliminates several models available in the literature, including the ones used in our latest {\sc starlight}-based papers on the stellar populations of CALIFA galaxies (e.g. Gonz\'alez Delgado et al.\  2015). Cid Fernandes et al.\ (2014) present a comparative study of optical spectral synthesis results obtained with different spectral bases, including the one used in this paper.

Dust effects are modelled as due to a single effective screen of V-band optical depth $\tau$, and a Calzetti et al.\ (2000) reddening law. The code does allow one to waive the assumption of a single $\tau$ for all populations, but for the sake of simplicity we stick to this assumption in this paper, postponing the exploration of multiple-$\tau$ fits to a forthcoming study where recombination emission lines are added as a further constraint on the analysis (Vale Asari et al., in prep.).

\section{Simulations}
\label{sec:Simulations}

As a first test of the new {\sc starlight} outlined above we carried out a set of controlled experiments whereby the observables of a theoretical galaxy are fitted. The simulations are designed analogously to those in Cid Fernandes et al.\ (2005), where the original {\sc starlight} was first tested. The goals here are twofold: {\em (i)} To gauge the performance of the code under different levels of signal-to-noise ($S/N$), and {\em (ii)} to evaluate the practical benefits of a joint UV photometry plus optical spectral analysis in comparison to a  purely optical one.

\subsection{Test galaxies}

We generate  test galaxies using the parameters (essentially $\vec{x}$ and $\tau$) obtained from the analysis of 260  CALIFA galaxies ranging from early to late types, thus spanning a range of physical properties (see Section \ref{sec:CALIFAGALEXsample} and Fig.\ \ref{fig:CMD_NUV_r}). This strategy has the advantage of ensuring that our test galaxies are both diverse and realistic, while also saving us the trouble of inventing test galaxies with {\it ad hoc} descriptions of the SFH.

The observables for these test galaxies, namely the 3700--6800 \AA\ optical spectrum $O_\lambda$ and the 
$m^{\rm obs}_l$ (where $l =$ NUV, FUV) magnitudes, were generated from their full synthetic spectra and then perturbed according to 

\begin{equation}
O_{\lambda} = O_{\lambda}^0 \left(1 + \frac{ {\cal N}(0,1) }{ S/N } \right)
\end{equation}
\begin{equation}
m^{\rm obs}_l = m^{0}_l +( 2.5\log e) \, \frac{ {\cal N}(0,1) }{ S/N }
\end{equation}

\noindent where $O_{\lambda}^0$ and $m_l^0$  are the original input spectrum and magnitudes, and
${\cal N}(0,1)$ is a gaussian deviate of zero mean and unit variance. Five levels of noise were considered: $S/N = 5$, 10, 20, 50, 100. (The corresponding errors in magnitudes are 0.217, 0.109, 0.054, 0.022 and 0.011, respectively.) Five realizations of the noise were made for each $S/N$. In total, the test sample consists of $260 \times 5 \times 5 = 6500$ galaxies.

To emulate actual fits inasmuch as possible we masked regions around the main emission lines, 
including the whole Balmer series up to H$\epsilon$. As in actual spectral fits (e.g. Gonz\'alez Delgado et al.\ 2015) the NaI D doublet was also masked because of its interstellar component.

Each version of each test galaxy was fitted twice: with and without the UV information. Fits considering only the optical spectrum are hereafter called OPT (for optical) fits, while those which also fit NUV and FUV data are called PHO (for photometry) fits.

\subsection{Example fits}
\label{sec:ExampleFits}

%***FIG***FIG***FIG***FIG***FIG***FIG***FIG***FIG***FIG***FIG***FIG***FIG***FIG***FIG***FIG***FIG
\begin{figure}
\includegraphics[width=0.5\textwidth]{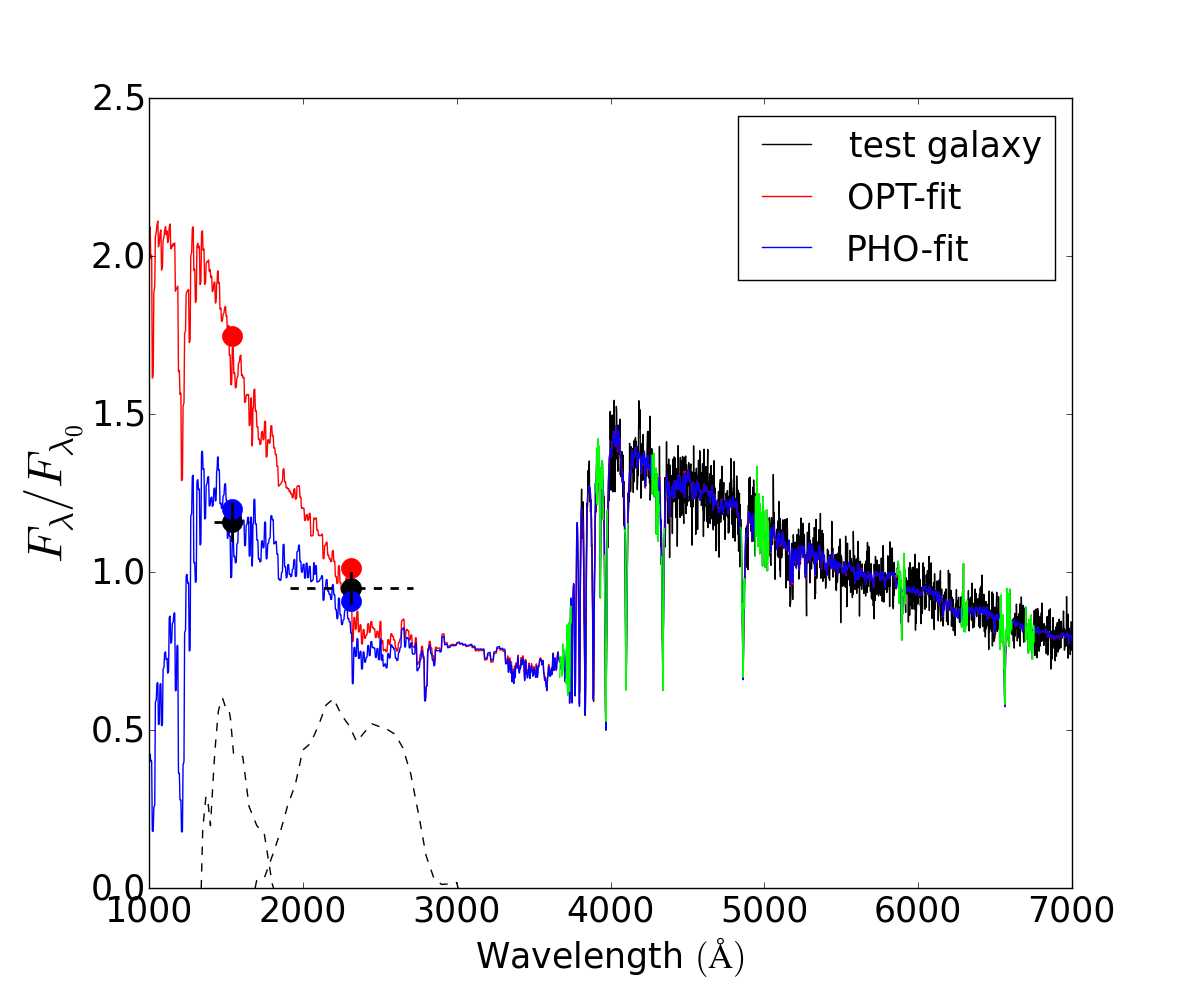}
\caption{Example optical-only (OPT) and optical+UV (PHO) fits of a simulated galaxy. The black line shows the observed spectrum, with masked regions marked in green. Black diamonds and error bars show the NUV and FUV fluxes---the corresponding filter transmission curves are shown as dashed black lines.
Red and blue lines show OPT and PHO fits, respectively. The two fits are indistinguishable in the optical, but diverge in the UV, where the lines are drawn as dashed to indicate that we do not have actual UV spectra, but only the photometry, indicated by filled circles.
The OPT fit overpredicts the UV fluxes (especially FUV) because of optically-insignificant but UV-dominant young populations (see text).
}
\label{fig:fit_example}
\end{figure}
%***FIG***FIG***FIG***FIG***FIG***FIG***FIG***FIG***FIG***FIG***FIG***FIG***FIG***FIG***FIG***FIG

%***FIG***FIG***FIG***FIG***FIG***FIG***FIG***FIG***FIG***FIG***FIG***FIG***FIG***FIG***FIG***FIG
\begin{figure}
\includegraphics[width=0.5\textwidth]{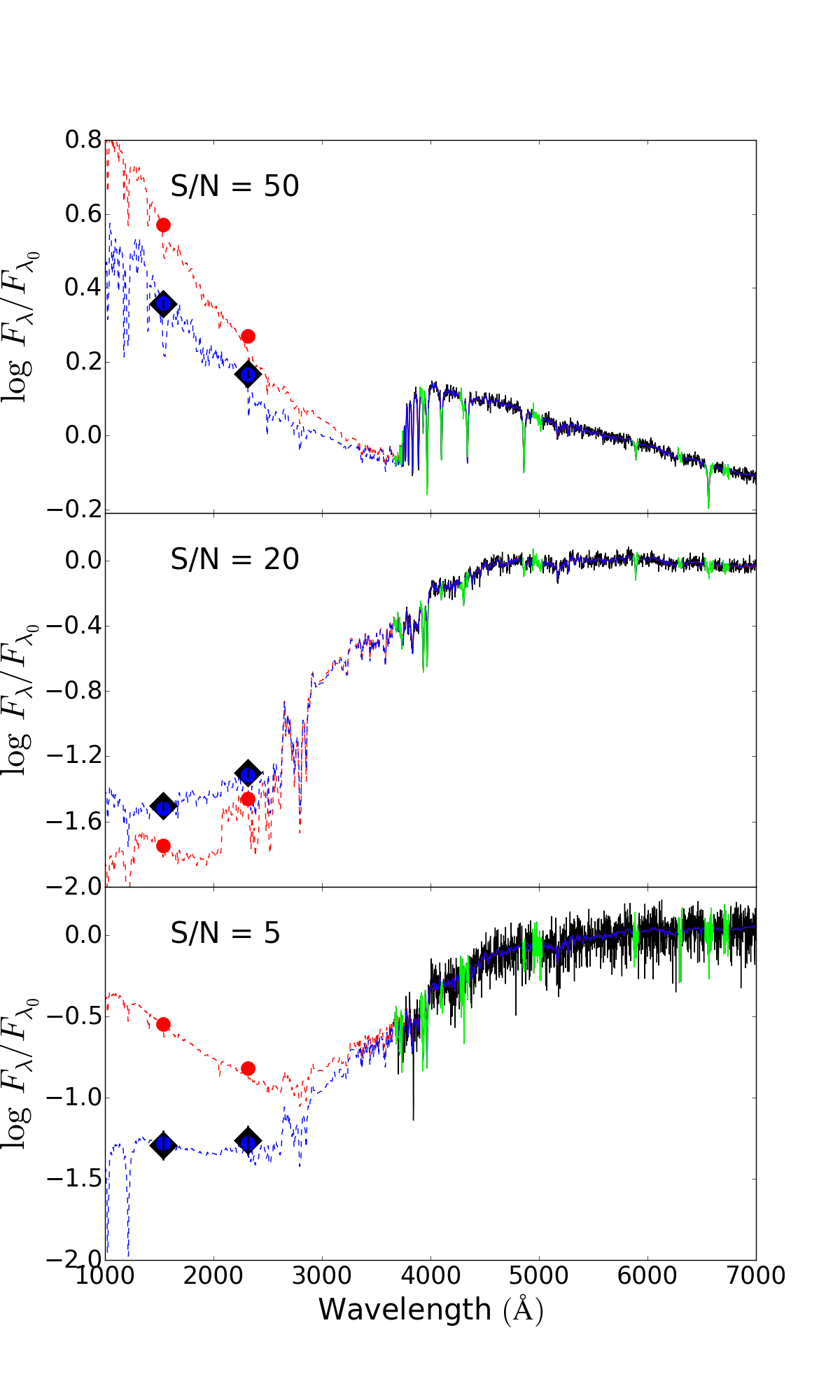}
\caption{
As Fig. \ref{fig:fit_example}  for three further examples of OPT and PHO fits to simulated galaxies. The three panels are representative of blue cloud (top panel), green valley (middle), and red sequence (bottom) galaxies. The labels indicate the $S/N$ of the simulated data.
}
\label{fig:MoreFitExamples}
\end{figure}
%***FIG***FIG***FIG***FIG***FIG***FIG***FIG***FIG***FIG***FIG***FIG***FIG***FIG***FIG***FIG***FIG

Fig.\ \ref{fig:fit_example} shows an example. The black spectrum and black diamond NUV and FUV points represent the input data for a test galaxy. The blue and red spectra are the results of PHO and OPT fits, respectively, and similarly for the filled blue and open red UV fluxes. Though the full  $\lambda$-by-$\lambda$ predicted UV spectrum is drawn, only the NUV and FUV fluxes are relevant in our analysis. 

The OPT and PHO fits are so similar in the optical that they cannot be told apart in Fig.\ \ref{fig:fit_example}. Indeed their $M_\lambda$ fluxes differ by less than 1\% on average over the fitted range (3700--6800 \AA). In the UV, however, they diverge. While the PHO fit matches the UV data to within the errors,  the OPT fit overpredicts the UV fluxes, specially in the FUV filter. That PHO fits perform better in the UV is of course not surprising, since they are designed to take the UV photometry into consideration, while OPT fits ignore it. What is perhaps unexpected is that the two fits so different in the UV yield nearly identical optical spectra.

This happens because the OPT fit ascribes 5\% of the light at $\lambda_0 = 5635$ \AA\ (our chosen reference wavelength) to populations of 30 Myr or younger. This small number reflects the insignificant contribution of these populations to the optical spectrum. Removing this component or replacing it by another one would make little difference for the optical fit. Yet, this same component overwhelms all the others at UV wavelengths (e.g. Kaviraj et al. 2007b). In the absence of UV constraints, {\sc starlight} sees no harm in depositing some small amount of light in this  population. Once it is informed about the UV fluxes, however, it realizes that some other combination of base elements must be sought to accommodate both the optical spectrum and the UV photometry. In the case at hand, the 1--30 Myr populations found in the OPT fit shrink to 2\% in the PHO fit, being replaced by an increase in populations of 30--100 Myr. 
In rough terms, one can summarize the change as a shift from populations of O and B to one of B and A stars.

At this point it is fit to open a parenthesis to mention the potential effects of differential extinction. Another way to achieve the same overall result obtained with the PHO fit in Fig.\ \ref{fig:fit_example} would be to attribute an extra reddening to the young population, an astrophysically attractive  solution given that young stars (and their surrounding nebulae) are known to suffer more extinction than the general stellar population (Calzetti, Kinney \& Storchi-Bergmann 1994; Charlot \& Fall 2000). Mathematically, accounting for this complication in our modelling would require changing equation \ref{eq:ModelFlux} and transforming our single dust parameter $\tau_V$ into an age-dependent vector. Though {\sc starlight} is prepared to handle multiple extinctions, experiments show that it does so much more reliably when information on recombination emission line fluxes (e.g. H$\alpha$ and H$\beta$) is modelled along with other observables. For clarity and simplicity, we postpone the presentation of this further (and qualitatively different) upgrade of the code to a future communication, noting that, because age-dependent dust optical depths alter the UV/optical balance, it might impact the results reported in this paper.

Going back to PHO and OPT fits, Fig.\ \ref{fig:MoreFitExamples} shows three further examples. In all cases PHO fits do an excellent job in fitting both the optical spectrum and the UV photometry. The tendency of OPT fits to overshoot the UV fluxes is illustrated by the top and bottom examples, but the middle one shows that the opposite can also happen. As observed in Fig.\ \ref{fig:fit_example}, the optical spectra are practically indistinguishable between PHO and OPT fits, indicating again that a small variation in the fraction of young stellar populations can be imperceptible in the optical spectra but produce an important change in the UV flux.

As a whole, these examples suggest that a simultaneous analysis of  optical spectra and UV photometry should bring some improvement in the estimation of the strength of young stellar populations in a galaxy. For a more global mapping of what actually changes from OPT to PHO let us examine the results of the full set of simulations.

\subsection{Input versus output: UV fluxes}

%***FIG***FIG***FIG***FIG***FIG***FIG***FIG***FIG***FIG***FIG***FIG***FIG***FIG***FIG***FIG***FIG
\begin{figure}
\includegraphics[width=0.5\textwidth]{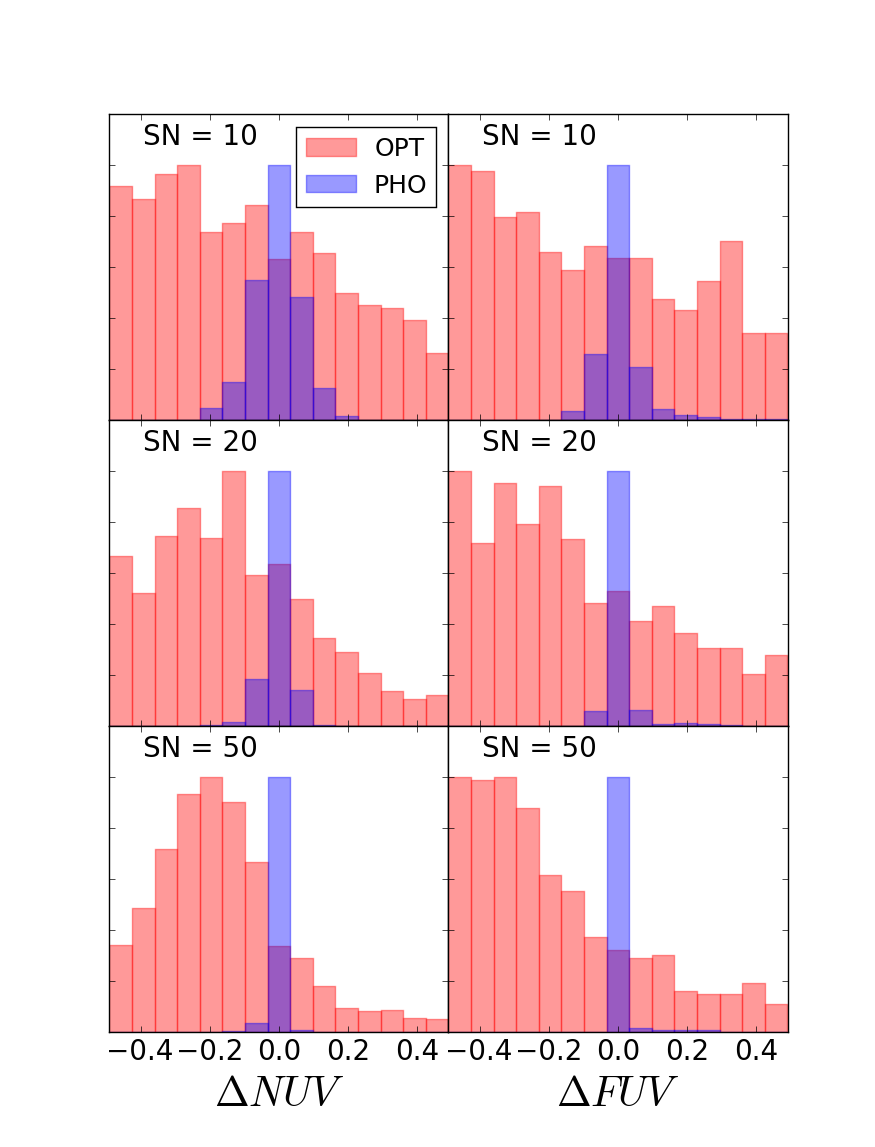}
\caption{Comparison between observed and predicted $NUV$ (left) and $FUV$ (right) magnitudes for simulations with $S/N = 10$, 20 and 50 (from top to bottom). In all panels $\Delta$ = Predicted $-$ Observed. Histograms have been scaled to the same peak.}
\label{fig:NUV_FUV_magnitudes}
\end{figure}
%***FIG***FIG***FIG***FIG***FIG***FIG***FIG***FIG***FIG***FIG***FIG***FIG***FIG***FIG***FIG***FIG

%***TAB***TAB***TAB***TAB***TAB***TAB***TAB***TAB***TAB***TAB***TAB***TAB***TAB***TAB***TAB
\begin{table*}
\centering % used for centering table
\begin{tabular}{c r r r r r} % centered columns (4 columns)
\multicolumn{6}{|c|}{Summary of simulations: $\overline{\Delta}\pm \sigma_{\Delta}$} \\
\hline %inserts double horizontal lines
Property (OPT) & \multicolumn{1}{|c|}{S/N = 5} & \multicolumn{1}{|c|}{S/N = 10} & \multicolumn{1}{|c|}{S/N = 20} & \multicolumn{1}{|c|}{S/N = 50} & \multicolumn{1}{|c|}{S/N = 100}\\ %[0.5ex] % inserts table
%heading
\hline % inserts single horizontal line
%r & 0.000 \pm 0.008 & 0.000 \pm 0.004 & 0.000 \pm 0.002 & 0.000 \pm 0.000 & 
%0.000 \pm 0.000 \\
NUV  &  $-0.25 \pm 0.64$  & $ -0.26 \pm 0.47 $ & $ -0.22 \pm 0.35 $ & $ -0.19 \pm 0.23 $ & $ -0.16 \pm 0.18  $\\
FUV  & $ -0.28 \pm 1.23 $ & $ -0.35 \pm 0.95 $ &$  -0.32 \pm 0.75 $ & $ -0.32 \pm 0.56 $ & $ -0.26 \pm 0.47 $ \\
%$\chi^2_ {\rm SPEC}$  & $1360.50 \pm 56.57 $ & $ 1361.07 \pm 55.07 $ & $ 1355.96 \pm %56.61 $&  $1355.42 \pm 54.01 $&   $1361.36 \pm 58.47$\\
$\chi^2_ {\rm SPEC}  / N_\lambda^{\rm eff}$ & $1.06 \pm 0.04$ & $1.06 \pm 0.04$ & 
$1.05 \pm 0.04 $& $1.05 \pm 0.04 $& $1.06 \pm 0.05$ \\
%adev ($\%$) & $17.38 \pm 0.46 $& $8.09 \pm 0.18 $& $3.98 \pm 0.09$ & 
%$1.59 \pm 0.03$ & $0.79 \pm 0.02 $\\
\hline %inserts single line
Property (PHO) & \multicolumn{1}{|c|}{S/N = 5} & \multicolumn{1}{|c|}{S/N = 10} & \multicolumn{1}{|c|}{S/N = 20} & \multicolumn{1}{|c|}{S/N = 50} & \multicolumn{1}{|c|}{S/N = 100}\\
\hline %inserts single line
%r & 0.000 \pm 0.008 & 0.000 \pm 0.004 & 0.000 \pm 0.002 & 0.000 \pm 0.000 & 
%0.000 \pm 0.000 \\
NUV  & $ 0.00 \pm 0.14  $& $ 0.00 \pm 0.08 $ & $ -0.00 \pm 0.04 $ &$  -0.00 \pm 0.02 $ & $ -0.00 \pm 0.02  $\\
FUV  & $ -0.00 \pm 0.21 $ & $ 0.00 \pm 0.12 $ & $ 0.01 \pm 0.08  $& $ 0.01 \pm 0.06 $ & $ 0.01 \pm 0.05 $ \\
%$\chi^2_ {\rm SPEC}$ & $1361.28 \pm 56.63 $ &  $1362.39 \pm 55.21 $ & $ 1358.87 \pm 57.18 $&  $1363.04 \pm 63.31$ &  $1380.87 \pm 59.4$\\
$\chi^2_ {\rm SPEC} / N_\lambda^{\rm eff}$ & $1.06 \pm 0.04$ &$ 1.06 \pm 0.04$ & 
$1.06 \pm 0.04 $& $1.06 \pm 0.05 $& $1.07 \pm 0.06 $\\
%adev ($\%$) & $17.38 \pm 0.47$ &$ 8.10 \pm 0.18$ &$ 3.98 \pm 0.09 $& $1.59 \pm 0.04$ & $0.80 \pm 0.03$ \\

\hline %inserts single line
\end{tabular}
\centering
\caption{Statistics of the simulations. For each of the NUV and FUV magnitudes the table lists the mean predicted minus observed difference ($\overline{\Delta}$) and its standard deviation ($\sigma_{\Delta}$). Also listed is $\chi^2_{\rm SPEC}/N_{\lambda}^{\rm eff}$,  a measure of the qualify of the fit of the optical spectrum. Columns are for the different levels of noise ($S/N = 5$, 10, 20, 50, 100).
}
\label{table:mag_prop} 
\end{table*}
%***TAB***TAB***TAB***TAB***TAB***TAB***TAB***TAB***TAB***TAB***TAB***TAB***TAB***TAB***TAB

To quantify the performance of the fits let us define $\Delta$ as the output $-$ input difference in some quantity (say, the NUV magnitude) and examine its statistics. Fig.\ \ref{fig:NUV_FUV_magnitudes} shows the histograms of $\Delta$ for both NUV and FUV magnitudes. Different panels are for different $S/N$ values, and blue and red lines are for PHO and OPT fits, respectively. 

Fig.\ \ref{fig:NUV_FUV_magnitudes}  reinforces the conclusions that: (1) PHO fits do match the UV fluxes, as designed to, and (2)  OPT fits are lousy predictors of UV fluxes, with large $\Delta m_{\rm NUV}$ and $\Delta m_{\rm FUV}$ even for high quality data. Table \ref{table:mag_prop} summarizes these results by listing the mean ($\overline{\Delta}$) and standard deviation ($\sigma_\Delta$) of $\Delta$ for $m_{\rm NUV}$  and $m_{\rm FUV}$ over all test galaxies. $\Delta$ and its dispersion behaves as expected for PHO fits, with $\overline{\Delta} \sim 0$ and $\sigma$ close to the expected noise levels. OPT fits, on the other hand, have a  tendency to overpredict the UV fluxes ($=$ underpredict  magnitudes, so $\Delta < 0$) by 0.1--0.3 mag, even for high quality data, and $\sigma_\Delta$ is well above the photometric errors. 

Table \ref{table:mag_prop} also lists the statistics of the $\chi2_{\rm SPEC}/N_{\lambda}^{\rm eff}$ figure of merit. The results show that the large differences in UV predictions between OPT and PHO fits occur for  equivalent performances insofar as the optical spectrum is concerned, corroborating the visual impression drawn from the examples in Figs.\ \ref{fig:fit_example} and  \ref{fig:MoreFitExamples}.

 \subsection{Input versus output: Physical properties}

%***TAB***TAB***TAB***TAB***TAB***TAB***TAB***TAB***TAB***TAB***TAB***TAB***TAB***TAB***TAB
\begin{table*}
\centering % used for centering table
\begin{tabular}{c r r r r r} % centered columns (4 columns)
\multicolumn{6}{|c|}{Summary of simulations: $\overline{\Delta}\pm \sigma_{\Delta}$} \\
\hline %inserts double horizontal lines
Property (OPT) & \multicolumn{1}{|c|}{S/N = 5} & \multicolumn{1}{|c|}{S/N = 10} & \multicolumn{1}{|c|}{S/N = 20} & \multicolumn{1}{|c|}{S/N = 50} & \multicolumn{1}{|c|}{S/N = 100}\\ %[0.5ex] % inserts table
%heading
\hline % inserts single horizontal line
$\log\, M$  & $ -0.03 \pm 0.16 $ &  $-0.02 \pm 0.12  $&$  -0.02 \pm 0.09 $ & $ -0.01 \pm 0.07 $ & $ -0.01 \pm 0.06 $ \\
$\tau_{V}$  & $ 0.03 \pm 0.16$  & $ 0.02 \pm 0.10 $ & $ 0.02 \pm 0.06  $& $ 0.01 \pm 0.03 $ & $ 0.01 \pm 0.02$  \\
%\\
$\langle\log\, t\rangle_{L}$  & $ -0.11 \pm 0.28 $ & $ -0.08 \pm 0.22 $ &  $-0.06 \pm 0.18 $ & $ -0.06 \pm 0.13 $ & $ -0.04 \pm 0.11 $ \\
$\langle\log\, t\rangle_{M}$  & $ -0.04 \pm 0.25  $& $ -0.02 \pm 0.18 $ & $ 0.00 \pm 0.14 $ & $ 0.00 \pm 0.11 $ &  $0.00 \pm 0.09 $ \\
$\langle\log\, Z\rangle_{L}$  & $ -0.06 \pm 0.31 $ & $ -0.05 \pm 0.22  $& $ -0.05 \pm 0.14 $ & $ -0.04 \pm 0.09  $& $ -0.04 \pm 0.06  $\\
$\langle\log\, Z\rangle_{M}$  &  $0.06 \pm 0.43 $ &$  0.04 \pm 0.33$  & $ 0.04 \pm 0.22 $ & $ 0.03 \pm 0.13 $ & $ 0.03 \pm 0.09$  \\
%\\
$x_{Y}$  &  $1.92 \pm 8.09 $ & $ 1.38 \pm 5.80  $& $ 1.07 \pm 5.16 $ & $ 0.73 \pm 4.15 $ &  $0.42 \pm 3.52 $ \\
$x_{I}$  & $ 3.56 \pm 15.03 $ & $ 2.62 \pm 11.65 $ & $ 1.47 \pm 7.81 $ &  $1.55 \pm 5.34 $ & $ 1.4 \pm 4.51 $ \\
$x_{O}$  & $ -4.47 \pm 16.05 $ & $ -3.16 \pm 12.68 $ & $ -1.71 \pm 8.88 $ & $ -1.46 \pm 5.99$  & $ -1.04 \pm 4.76 $ \\
$\mu_{Y}$  &  $0.09 \pm 2.88 $ & $ -0.05 \pm 0.74 $ & $ -0.06 \pm 0.44 $ & $ -0.05 \pm 0.47 $ &$  -0.05 \pm 0.39 $ \\
$\mu_{I}$  &  $2.65 \pm 12.84 $ & $ 1.61 \pm 9.23 $ & $ 0.44 \pm 5.40 $ & $ 0.43 \pm 3.84 $ & $ 0.44 \pm 3.03 $ \\
$\mu_{O}$  & $ -2.74 \pm 13.27 $ & $ -1.56 \pm 9.49  $& $ -0.37 \pm 5.51 $ & $ -0.38 \pm 4.03 $ & $ -0.38 \pm 3.21 $ \\
%adds vertical space
\hline %inserts single line
Property (PHO) & \multicolumn{1}{|c|}{S/N = 5} & \multicolumn{1}{|c|}{S/N = 10} & \multicolumn{1}{|c|}{S/N = 20} & \multicolumn{1}{|c|}{S/N = 50} & \multicolumn{1}{|c|}{S/N = 100}\\
\hline %inserts single line
$\log\, M$  & $ 0.00 \pm 0.15 $ & $ 0.00 \pm 0.11 $ & $ 0.01 \pm 0.08 $ & $ 0.01 \pm 0.06 $ & $ 0.01 \pm 0.05  $\\
$\tau_{V}$  & $ 0.03 \pm 0.16  $& $ 0.01 \pm 0.10  $& $ 0.01 \pm 0.05 $ & $ 0.00 \pm 0.03 $ & $ 0.00 \pm 0.02 $ \\
%\\
$\langle\log\, t\rangle_{L}$  & $ -0.01 \pm 0.23  $& $ 0.01 \pm 0.16 $ &  $0.02 \pm 0.10 $ & $ 0.02 \pm 0.06 $ &  $0.02 \pm 0.05 $ \\
$\langle\log\, t\rangle_{M}$  &  $-0.01 \pm 0.23$  & $ 0.00 \pm 0.17 $ & $ 0.02 \pm 0.13 $ & $ 0.02 \pm 0.10 $ & $ 0.03 \pm 0.09  $\\
$\langle\log\, Z\rangle_{L}$  & $ -0.05 \pm 0.30  $& $ -0.05 \pm 0.22  $& $ -0.03 \pm 0.13 $ & $ -0.03 \pm 0.07 $ & $ -0.01 \pm 0.05 $ \\
$\langle\log\, Z\rangle_{M}$  &  $-0.05 \pm 0.37 $ & $ -0.03 \pm 0.26  $& $ -0.03 \pm 0.15 $ & $ -0.02 \pm 0.10  $& $ -0.01 \pm 0.08 $ \\
%\\
$x_{Y}$  &  $-0.73 \pm 5.48 $ & $ -0.88 \pm 3.98  $& $ -0.86 \pm 2.98 $ & $ -0.50 \pm 2.08 $ & $ -0.56 \pm 1.75 $ \\
$x_{I}$  & $ 3.92 \pm 14.76 $ &  $2.00 \pm 10.37 $ & $ 0.69 \pm 7.15 $ & $ 0.51 \pm 4.45 $ & $ 0.59 \pm 3.81 $ \\
$x_{O}$  & $ -2.16 \pm 15.24 $ & $ -0.29 \pm 10.95 $ &  $0.99 \pm 7.24 $ &  $0.79 \pm 4.41 $ &$  0.75 \pm 3.53 $ \\
$\mu_{Y}$  & $ 0.08 \pm 3.58  $& $ -0.06 \pm 0.52 $ & $ -0.08 \pm 0.33 $ & $ -0.05 \pm 0.27 $ & $ -0.06 \pm 0.20 $ \\
$\mu_{I}$  &  $1.67 \pm 11.15  $& $ 0.44 \pm 7.32 $ & $ -0.41 \pm 4.50 $ & $ -0.36 \pm 3.00 $ & $ -0.32 \pm 2.43 $ \\
$\mu_{O}$  & $ -1.74 \pm 11.65 $ & $ -0.38 \pm 7.59 $ & $ 0.49 \pm 4.59  $& $ 0.41 \pm 3.09 $ & $ 0.38 \pm 2.44  $\\
%adds vertical space
\hline %inserts single line
\end{tabular}
\centering
\caption{Statistics of the simulations. For each physical property the table lists the mean simulated minus original difference ($\overline{\Delta}$) and its standard deviation ($\sigma_{\Delta}$) for $S/N$ varying from 5 to 100. The age-grouped light ($x$) and mass ($\mu$) fractions  are given in percentage.
}
\label{table:phy_prop}
\end{table*}
%***TAB***TAB***TAB***TAB***TAB***TAB***TAB***TAB***TAB***TAB***TAB***TAB***TAB***TAB***TAB

The OPT $\times$ PHO comparisons above were carried out in a space of observable quantities (UV magnitudes and optical spectral residuals). We now compare OPT and PHO in terms of physical properties. Table \ref{table:phy_prop}  (built to be similar to Table 1 in Cid Fernandes et al.\ 2005) lists the statistics of $\Delta$ as a function of $S/N$ as obtained with these two approaches,  and for the suite of properties discussed next.

\subsubsection{Stellar mass and extinction}

Stellar mass ($M_\star$) is not an explicit parameter in {\sc starlight}, but a byproduct of the light fraction population vector ($\vec{x}$) translated to mass fractions ($\vec{\mu}$) through the known light-to-mass ratios of the $N_\star$ base populations. As widely known (e.g. Salim et al.\ 2007; Taylor et al.\ 2011), and foregoing IMF-related uncertainties, $M_\star$ is a relatively robust quantity in both photometric and spectroscopic analysis.

Table \ref{table:phy_prop} shows that $\log M_\star$ is recovered very accurately in the simulations, with $\overline{\Delta} \sim 0$ for both PHO and OPT fits and any $S/N$. The dispersion (i.e.\ the uncertainty) in $\Delta \log M_\star$ for OPT fits ranges from $\sigma_\Delta = 0.16$ to 0.06 dex from $S/N$ between 5 and 100, and very slightly smaller for PHO fits. UV info therefore does not help constrain $M_\star$ in any significant way. 

The same happens with the dust parameter $\tau_{V}$. In this case, for $S/N=20$  we have $\overline{\Delta}\pm\sigma_{\Delta} = 0.02\pm 0.06$  for OPT-fits,  and 
$ 0.01\pm 0.05$  for PHO-fits. In both cases $\tau_{V}$ is recovered to a similar level of precision.
(We expect  UV constraints to play a more relevant role in the multiple-$\tau_V$, differential extinction {\sc starlight} modelling mentioned in Section \ref{sec:ExampleFits}.)

\subsubsection{Mean stellar age and metallicity}

Table \ref{table:phy_prop} further lists the $\overline{\Delta} \pm \sigma_\Delta$ values for the luminosity weighted mean (log) age ($\langle\log t \rangle_L$) and metallicity ($\langle\log Z \rangle_L$). The first moment of the age distributions is given by $\langle\log t\rangle_L \equiv \sum_j x_j \times \log t_j$, and similarly for the metallicity. Mass weighted versions of these quantities are obtained replacing the light-fraction population vector $\vec{x}$ by its mass-fraction counterpart $\vec{\mu}$.

Examining the entries for these age and metallicity moments in   Table \ref{table:phy_prop} we see that $\sigma_\Delta$ decreases systematically from OPT to PHO fits, particularly for $\langle\log t \rangle_L$. For $S/N = 20$, for example, the addition of GALEX information to the optical spectrum brings the dispersion in $\Delta \langle\log t \rangle_L$ from 0.18 to 0.10 dex.
OPT fits also tend to be slightly biased towards younger ages, by $\overline{\Delta} \langle\log t \rangle_L = -0.04$ to $-0.11$ dex as $S/N$ decreases from 100 to 5, whereas PHO fits are not biased for any $S/N$.
%The bias is smaller for the mass weighted mean age $\langle\log t \rangle_M$ than for $\langle\log t \rangle_L$, indicating that this difference derives from differences in the young populations (which contribute much more in light than in mass), as previously seen in the   discussion of the examples in Figs.\ \ref{fig:fit_example}  and \ref{fig:MoreFitExamples}.
The bias is smaller for the mass weighted mean age $\langle\log t \rangle_M$ than for $\langle\log t \rangle_L$, indicating that this difference derives from differences in the young populations, which contribute much more in light than in mass. 

The simulations indicate that the GALEX input also improves the stellar metallicity estimates, e.g. from $0.04\pm 0.22$ to $-0.03\pm 0.15$ dex in $\Delta \langle \log Z \rangle_M$ for OPT and PHO $S/N = 20$ fits respectively. As for the mean stellar age, this improvement reflects the enhancement in the ability to estimate the contribution of young stellar populations.

\subsubsection{Star formation history: Condensed population vector}

There are several ways to process {\sc starlight}'s output population vectors into quantities which describe a galaxy's SFH. One which has been widely explored in the past (Gonz\'alez Delgado et al.\ 2004; Cid Fernandes et al.\ 2004; Cid Fernandes et al.\ 2005) is to rebin $\vec{x}$ on to `young' ($t_{j} < 10^{8}$ yr), `intermediate-age' ($10^{8}\leq t_{j}\leq 10^{9}$ yr), and `old' ($t_{j} > 10^{9}$ yr) populations ($x_{Y}$, $x_{I}$ and $x_{O}$, respectively). 

Table \ref{table:phy_prop} shows the statistics of $\Delta x_Y$, $\Delta x_I$, $\Delta x_O$. As expected, the  $\sigma_\Delta$ dispersions decrease for increasing $S/N$. More relevantly to this paper, all components of this condensed population vector have smaller uncertainties when UV constraints are added, i.e. as one goes from OPT to PHO fits. The gain is markedly larger for the youngest components, as might be expected given that young populations, when present, have a dominant contribution in the UV. Focusing again on the $S/N = 20$ simulations, we obtain that $\sigma_{\Delta x_Y}$ decreasing by $\sim 42\%$ as UV constraints are incorporated, while $\sigma_{\Delta x_I}$ and $\sigma_{\Delta x_O}$ decrease by some 8 and 18\% respectively. The same conclusion applies to the condensed mass-fractions population vector $(\mu_Y,\mu_I,\mu_O)$, also included in Table \ref{table:phy_prop}. 

In line with the mean age and metallicity results reported just above, we conclude that the simulations corroborate the basic intuitive notion that the addition of UV information to an optical spectral analysis is specially helpful in constraining the properties of young stellar populations (up to $\sim 300$ Myr). It is therefore natural to expect our optical+UV {\sc starlight} analysis to be particularly relevant for systems containing O, B and/or A stars, the UV-dominant component under most circumstances. This expectation is born out in the next Section.

\section{Application to CALIFA+GALEX data}
\label{sec:sample}

The experiments above served to validate the new capabilities of {\sc starlight}, as well as to provide a general sense on the changes resulting from the addition of UV photometry to an optical spectrum as observables in the analysis. In this section we present an application to real data, combining CALIFA spectra with GALEX magnitudes. 
%This being the first application of of the new code, the gist  is mainly on what changes from OPT to PHO fits. 

\subsection{Data and sample}

\subsubsection{CALIFA}

The Calar Alto Legacy Integral Field Area survey,  first described  by S\'anchez et al.\ (2012), has collected optical datacubes for over 500 galaxies, two hundred of which have already been made available in the 1$^{st}$  (Husemann et al.\ 2013) and 2$^{nd}$ (Garc\'ia Benito et al.\ 2015) data releases. Its targets are drawn from a diameter-selected sample fully characterized in Walcher et al.\ (2014). The data were collected at the 3.5m telescope of Calar Alto with  the Potsdam Multi-Aperture Spectrometer (Roth et al.\ 2005)  in the PPaK mode (Verheijen et al.\ 2004). 

Each galaxy is observed with two spectral settings which we combine to reduce the effects of vignetting on the data. This is done within version  1.5 of the  CALIFA pipeline. Our final datacubes cover about 1 arcmin$^2$ with  a spatial resolution of $\sim 3$ arcsec (FWHM) and cover the rest-frame spectral range from 3700 to 6800 with 6 \AA\ resolution. This rich data set has been used in over 30 published papers, some of which (P\'erez et al.\ 2013; Gonz\'alez Delgado et al.\ 2014, hereafter GD14) are based on results of a {\sc starlight} analysis of the optical spectra, i.e. `OPT fits' in the notation of this paper. Pre-processing steps (from Galactic extinction to Voronoi binning) and other details are discussed in Cid Fernandes et al.\ (2013).

\subsubsection{GALEX}

Nearly two thirds of the galaxies in the CALIFA mother sample have UV observations available from the Galaxy Evolution Explorer (GALEX) archive (Martin et al.\ 2005). In most cases these include both far-ultraviolet (FUV, effective wavelength $\lambda_{\rm eff}\sim$ 1542 \AA) and near-ultraviolet (NUV, $\lambda_{\rm eff}\sim$ 2274 \AA) bands. The data come from the GR6 data release. The GALEX archive provides simultaneous co-aligned FUV and NUV images with a field of view (FoV) of 1.2 degrees wide, spatial scale of 1.5 arcsec per pixel and a spatial resolution  of $\sim$ 4.5 arcsec (FWHM). 

For our sample the average uncertainties in the integrated data are 0.03 and 0.06 mag for NUV and FUV magnitudes respectively.
Because our sources are all nearby and relatively bright, these errors are about half the typical GALEX uncertainties.

\subsubsection{Combining CALIFA+GALEX data}
\label{sec:CALIFAGALEXsample}

%***FIG***FIG***FIG***FIG***FIG***FIG***FIG***FIG***FIG***FIG***FIG***FIG***FIG***FIG***FIG***FIG
\begin{figure}
\centering
\includegraphics[width=0.5\textwidth]{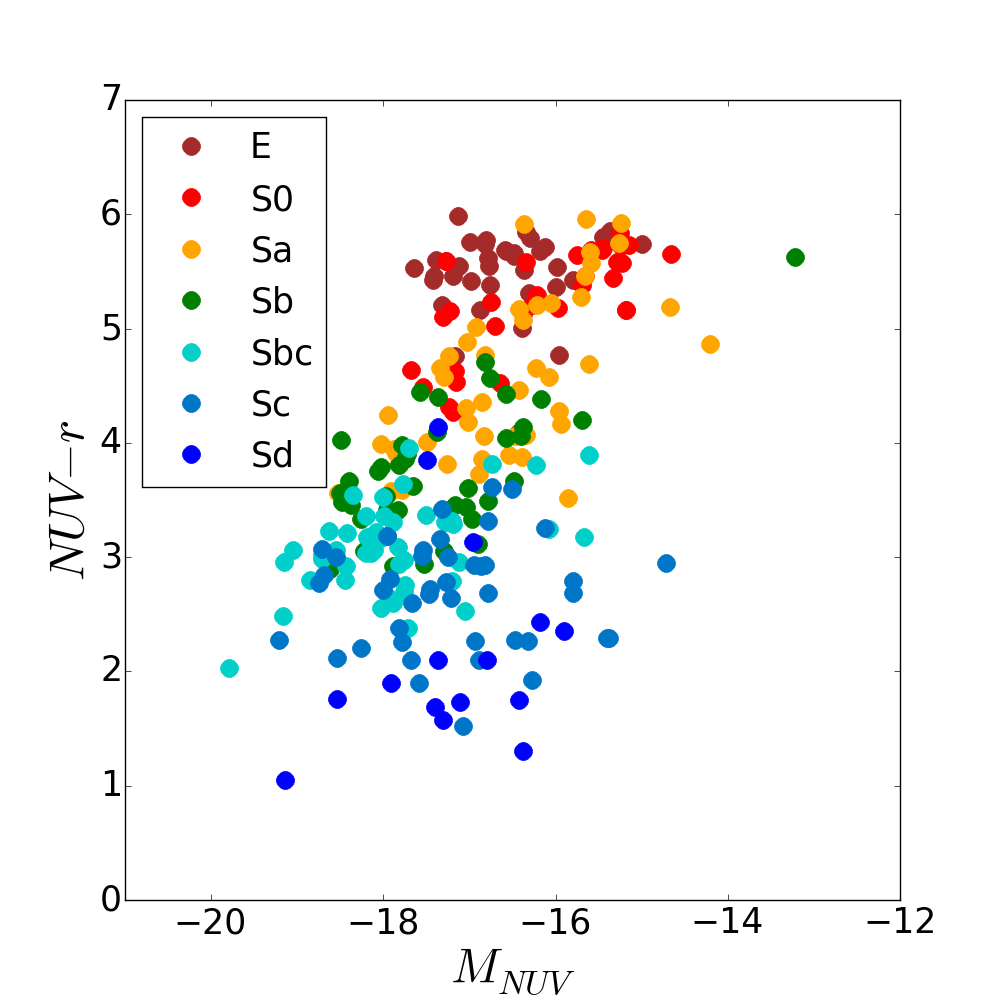}
\caption{Distribution of the 260 galaxies in the colour magnitude diagram $M_{NUV}$ vs $NUV-r$. Colour codes distinguish the morphological type, as labelled.  
}
\label{fig:CMD_NUV_r}
\end{figure}
%***FIG***FIG***FIG***FIG***FIG***FIG***FIG***FIG***FIG***FIG***FIG***FIG***FIG***FIG***FIG***FIG

Our combined CALIFA + GALEX sample contains 260 galaxies  ranging from early to late types, being a representative subset of the CALIFA sample as whole. This is illustrated in Fig.\ \ref{fig:CMD_NUV_r}, which shows galaxies scattered from the blue cloud to the red-sequence in a UV-optical colour magnitude diagram (CMD). Hubble types are coded by colours. Numerically, the sample contains 37 Ellipticals, 32 S0, 47 Sa, 41 Sb, 46 Sbc, 43 Sc, and 14 Sd (see Walcher et al.\ 2014 for the morphological composition of the CALIFA mother sample).

To combine our CALIFA datacubes with GALEX images we  use the MONTAGE software. We perform a resampling of the GALEX images to the same spatial scale as CALIFA. We then align and cut the GALEX images using the WCS to obtain processed FUV and NUV images with the same FoV of our CALIFA datacubes. Galactic extinction corrections following Wyder et al.\ (2007) were applied to the data.

These spatially matched data sets are ideal to circumvent the usual uncertainties associated with aperture effects. Indeed, this is the central motivation to use CALIFA data in this pilot study.

All the analysis in this section is based on spatially integrated data, obtained by collapsing the datacubes to a single optical spectrum and performing the UV photometry over the corresponding area. In Section \ref{sec:Discussion} we further explore results obtained from four spatial extractions: $r < 0.5$, $r < 1$, $1 < r  < 2$, and $r > 1$, where $r$ denotes the radial distance to the nucleus in units of the optical Half Light Radius (HLR, defined as in Cid Fernandes et al.\ 2013). None of our galaxies have a HLR smaller that the FWHM of the GALEX PSF, and most (247/260) have HLR larger than 8 arcsec, so these extractions are broad enough to avoid resolution issues.

\subsection{Results}

\label{sec:Results}

In analogy with the sequence followed  in Section \ref{sec:Simulations}, we first present the results regarding {\sc starlight}'s algorithmic goal, which is to fit the input observables (Sec.\  \ref{sec:Results_Fits}), and then in terms of the stellar population properties derived from the fits (Sec.\ \ref{sec:Results_PhysicalProperties}). Except for the observational errors, which in this case come from the actual data, the {\sc starlight} analysis was performed exactly as described in Section \ref{sec:Simulations}, with OPT fits analyzing only the 3700--6800 \AA\ spectra, and PHO fits adding the NUV and FUV magnitudes to the fit.

\subsubsection{{\sc starlight}  fits}
\label{sec:Results_Fits}

%***FIG***FIG***FIG***FIG***FIG***FIG***FIG***FIG***FIG***FIG***FIG***FIG***FIG***FIG***FIG***FIG
\begin{figure}
\includegraphics[width=0.5\textwidth]{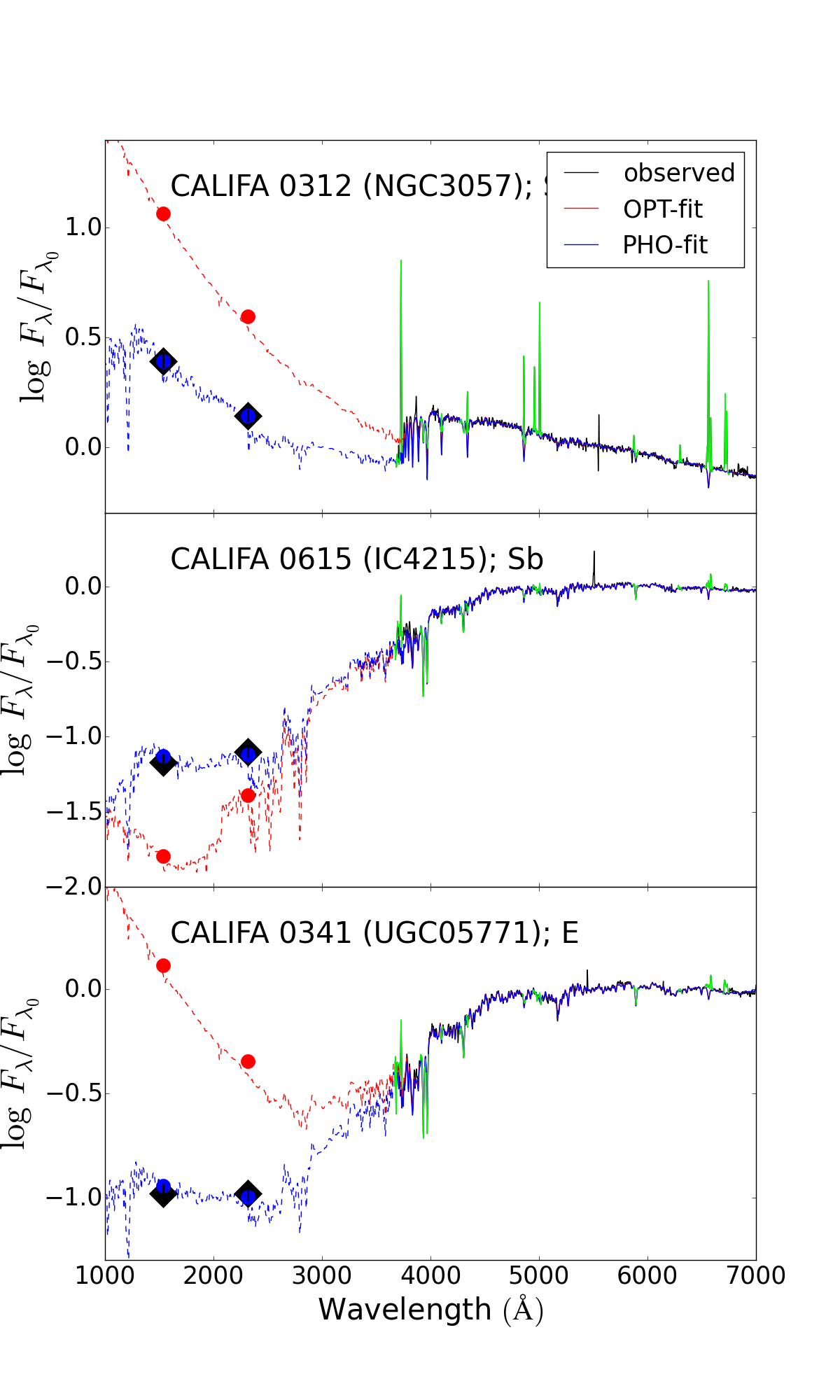}
\caption{Example OPT and PHO fits for three galaxies of different morphological types:  The Sd galaxy NGC 3057 (top),  the Sb IC 4215 (middle), and the elliptical UCG05771 (bottom).
}
\label{fig:RealFitsExamples}
\end{figure}
%***FIG***FIG***FIG***FIG***FIG***FIG***FIG***FIG***FIG***FIG***FIG***FIG***FIG***FIG***FIG***FIG

%***FIG***FIG***FIG***FIG***FIG***FIG***FIG***FIG***FIG***FIG***FIG***FIG***FIG***FIG***FIG***FIG
\begin{figure} 
\centering
\includegraphics[width=0.5\textwidth]{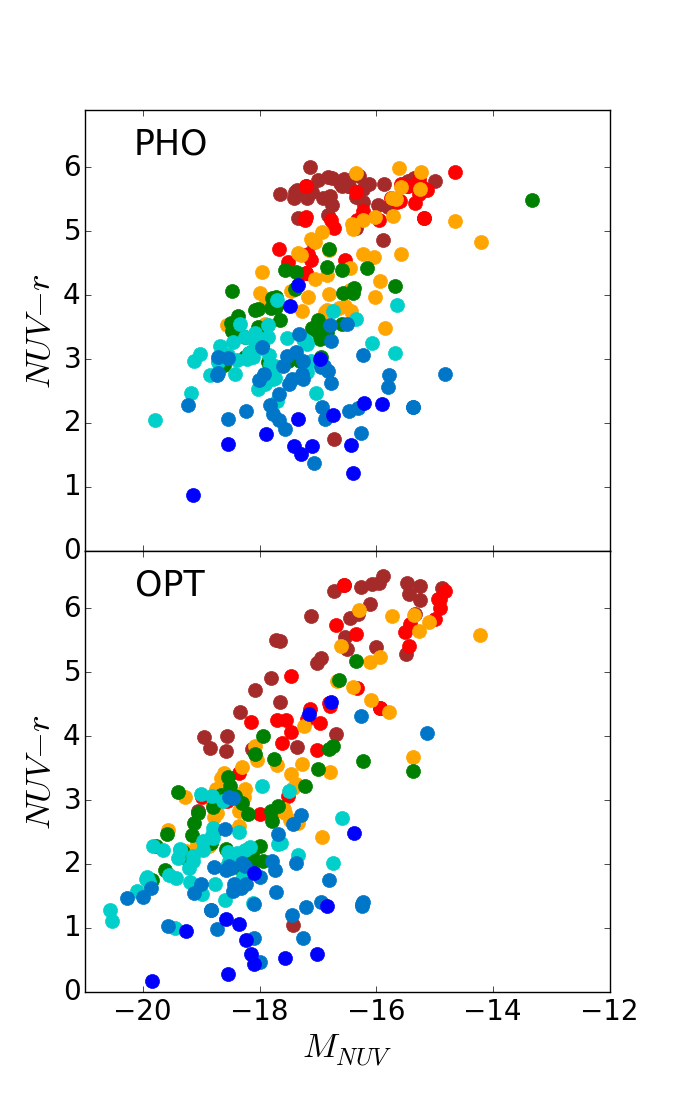}
\caption{Same as Fig.\ \ref{fig:CMD_NUV_r}  but for the synthetic magnitudes derived from PHO (top) and OPT (bottom) fits.
}
\label{fig:CMD_NUV_r_OPTxPHO}
\end{figure}
%***FIG***FIG***FIG***FIG***FIG***FIG***FIG***FIG***FIG***FIG***FIG***FIG***FIG***FIG***FIG***FIG

Fig.\ \ref{fig:RealFitsExamples} exemplifies the {\sc starlight} fits with three galaxies: NGC 3057 (top), IC 4215 (middle), and UCG 05771 (bottom). As in Fig.\ \ref{fig:MoreFitExamples}, OPT fits are shown in red and PHO fits in blue. As in the simulations, the optical spectra are equally well fitted in both kinds of fits. For instance, the mean percent deviation between $O_\lambda$ and $M_\lambda$ (eq.\ 6 in Cid Fernandes et al.\ 2013) are 2.8 and 3.1\%  in OPT and PHO fits. Also as in the simulations,  differences emerge in the UV. 
Again, UV fluxes tend to be overpredicted in OPT fits (top and bottom panels), but this is just a tendency, not a general rule. Cases like
IC 4215 (middle panel), where the OPT predictions fall short of the observed UV fluxes, also happen.

Fig.\ \ref{fig:CMD_NUV_r_OPTxPHO} shows the $NUV$ versus $NUV-r$  CMD derived from the synthetic photometry over the fitted spectra. As expected, the PHO-based  CMD (upper panel) matches the observed one (Fig.\ \ref{fig:CMD_NUV_r}), with rms differences of just  0.025 mag in $M_{NUV}$  and 0.044 in $NUV - r$. OPT fits, however, predict a wrongly shaped CMD. The incorrectly predicted $NUV$ fluxes produce shifts in both $M_{NUV}$ and $NUV - r$. The red sequence scatters towards both redder and bluer colours (as can be seen comparing the location of E galaxies in the two panels), while late type systems become both bluer and more luminous. As found in the simulations, OPT fits are poor predictors of the UV properties, particularly for systems in the blue cloud.

\subsubsection{Physical properties: Mass and dust attenuation}

\label{sec:Results_PhysicalProperties}

%***FIG***FIG***FIG***FIG***FIG***FIG***FIG***FIG***FIG***FIG***FIG***FIG***FIG***FIG***FIG***FIG
\begin{figure*}
\includegraphics[width=1\textwidth]{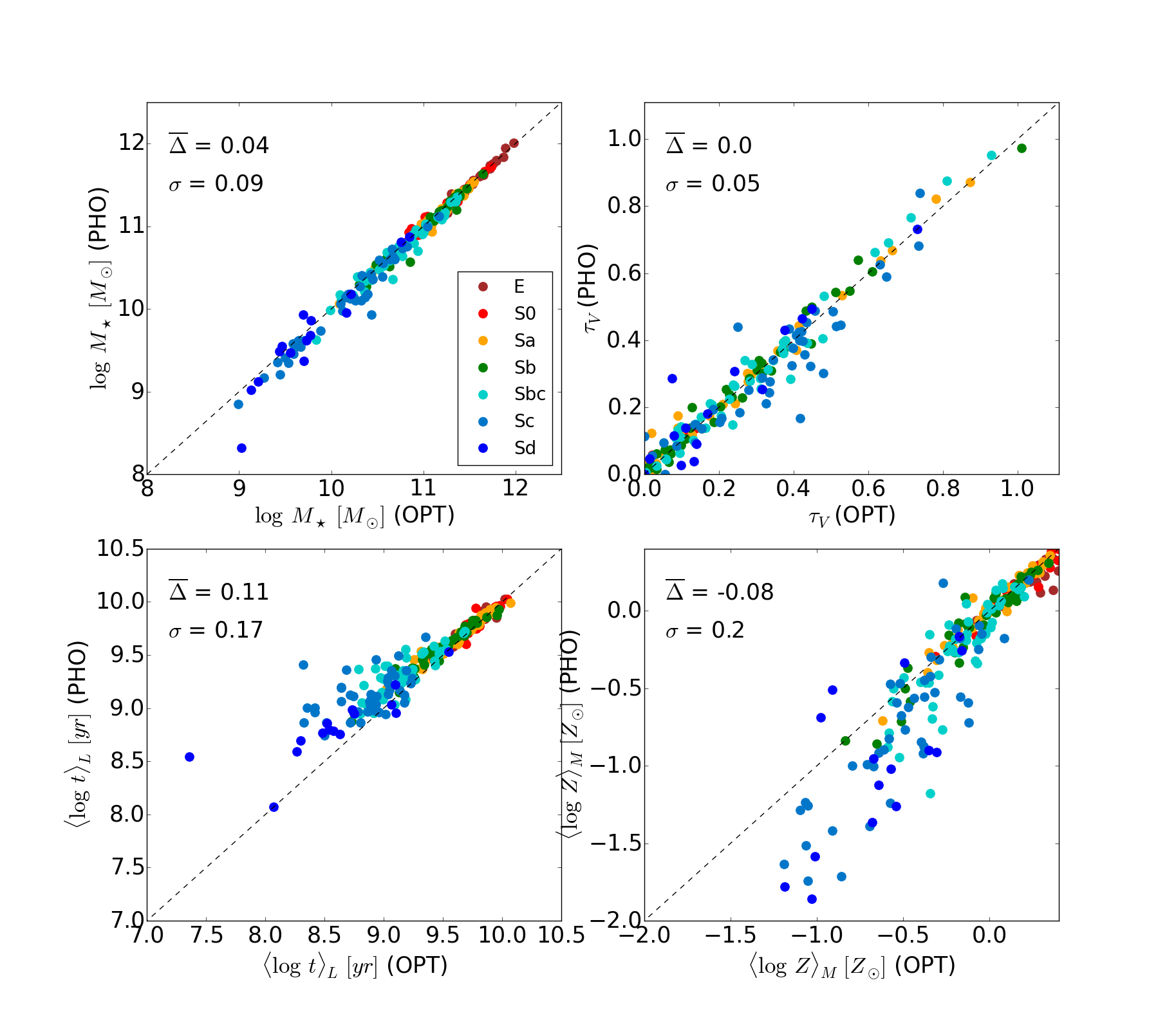}
\caption{Comparison of physical properties obtained with OPT and PHO fits. Each panel lists the average of the $\Delta = $ PHO $-$ OPT difference ($\overline{\Delta}$) and its  standard deviation ($\sigma_\Delta$). 
}
\label{fig:CALIFA_prop}
\end{figure*}
%***FIG***FIG***FIG***FIG***FIG***FIG***FIG***FIG***FIG***FIG***FIG***FIG***FIG***FIG***FIG***FIG

Let us now turn the focus from the observables to the stellar population properties derived from the analysis.

Fig.\ \ref{fig:CALIFA_prop} compares PHO and OPT results for a series of properties. As in Fig.\ \ref{fig:CMD_NUV_r}, points are colour coded by the Hubble type, an indirect but very efficient way of summarizing the properties of a galaxy, including its stellar populations (Kennicutt 1998; Gonz\'alez Delgado et al.\ 2015). 

Stellar masses, shown in the top-left panel, are essentially the same in PHO and OPT fits. Defining $\Delta$ as the PHO $-$ OPT difference in the value of any given property, we find  a mean value of $\overline{\Delta} = 0.04$ dex  for $\log M_\star$, with a dispersion $\sigma_\Delta = 0.09$ dex. The scatter is somewhat larger for late type systems, but still very small. For instance, for Sc--Sd galaxies (blue points) we obtain $\overline{\Delta}= 0.05$ and $\sigma_\Delta = 0.1$ dex, whereas for E-S0-Sa we obtain $\overline{\Delta}=0.01$ and $\sigma_\Delta = 0.05$ dex.

Regarding the $\tau_V$ values, we again observe no significant differences between PHO and OPT results, as seen in the top-right panel of Fig.\ \ref{fig:CALIFA_prop}. In this case we obtain $\overline{\Delta} = 0.0$ and $\sigma = 0.05$.
Late type systems (bluer points) are again the ones with a larger dispersion; $\sigma = 0.07$ if only Sc and Sd galaxies are considered. As a whole, however, and in agreement with the results of our simulations, neither $M_\star$ nor $\tau_V$ estimates gain much from the addition of UV constraints to a purely optical, conventional {\sc starlight} spectral fit.

\subsubsection{A note on why $\tau_V({\rm PHO}) \sim \tau_V({\rm OPT})$}

The apparently negligible  impact that UV information has upon our derived values of $\tau_V$ is perhaps surprising in light of the known sensitivity of UV fluxes to dust. Though subtle, the reasons for this somewhat counter intuitive result are easily understood. This section opens a parenthesis to explain them.

First, recall that OPT fits over predict the UV flux, so that PHO fits must find solutions which produce less UV. At the same time, and this is the key constraint here, PHO fits must keep the predicted optical spectrum essentially unchanged, since they must (by design) match both UV and optical data. UV fluxes can be diminished by: (1) increasing $\tau_V$, (2) decreasing the contribution of young stars ($x_Y$), and/or (3) increasing the age of the young population. The latter of these three (non-exclusive) alternatives is much more plausible than the others.

The first of these solutions is problematic. More dust may lead to the correct UV fluxes, but the larger reddening would then require increasing $x_Y$ to bluen the optical colors back to the observed values. More $x_Y$ would in turn imply more UV and require further increasing $\tau_V$ until an optimal balance is achieved. Note that even if this works in terms of colors, the increase in $x_Y$ would dilute absorption lines and hence degrade the quality of the optical fit. Increasing the age and metallicity of the old populations could perhaps restore the observed strengths of absorption features, but at this point it is clear that this route is a highly contrived one. 
The second alternative suggested above, namely, decreasing $x_Y$, sounds less problematic, but it is not a complete solution per se since the missing light must be replaced by something else. This `something else' should look like young stars in the optical but have a smaller UV per optical photon output. 

This brings us to the third and more natural solution: Aging the young population. From the optical point of view, populations of a few Myr or a few tens of Myr are very similar. For instance, and fixing $Z$ at  $Z_\odot$ for convenience, populations of 3 and 50 Myr have $g - r$ colors of $-0.5$ and $0$, and $D_n(4000)$ indices (Balogh et al.\ 1999) of 0.9, and 1.0, respectively. These relatively small differences contrast with the strong evolution of the UV, with $NUV - r$ changing from $\sim -1.3$ to $+0.5$ over the same time span. A change from populations of a few Myr to a few tens of Myr therefore produces the kind of result we need: less UV-per-optical emission at $\sim$ constant optical colors. As discussed later in Section \ref{sec:SFH}, the retrieved SFHs of OPT and PHO fits confirm this interpretation.

Aging the youngest populations thus offers the best way of simultaneously satisfying optical and UV constraints, which explains why $\tau_V$ remains approximately unaltered between OPT and PHO fits. It is adequate to recall that this conclusion applies to the simple foreground dust screen scenario adopted in this paper. Fits which allow for population-dependent $\tau_V$-values will certainly be more sensitive to the addition UV information.

\subsubsection{Physical properties: Mean stellar age and metallicity}

%***FIG***FIG***FIG***FIG***FIG***FIG***FIG***FIG***FIG***FIG***FIG***FIG***FIG***FIG***FIG***FIG
\begin{figure}
\centering
\includegraphics[width=0.45\textwidth]{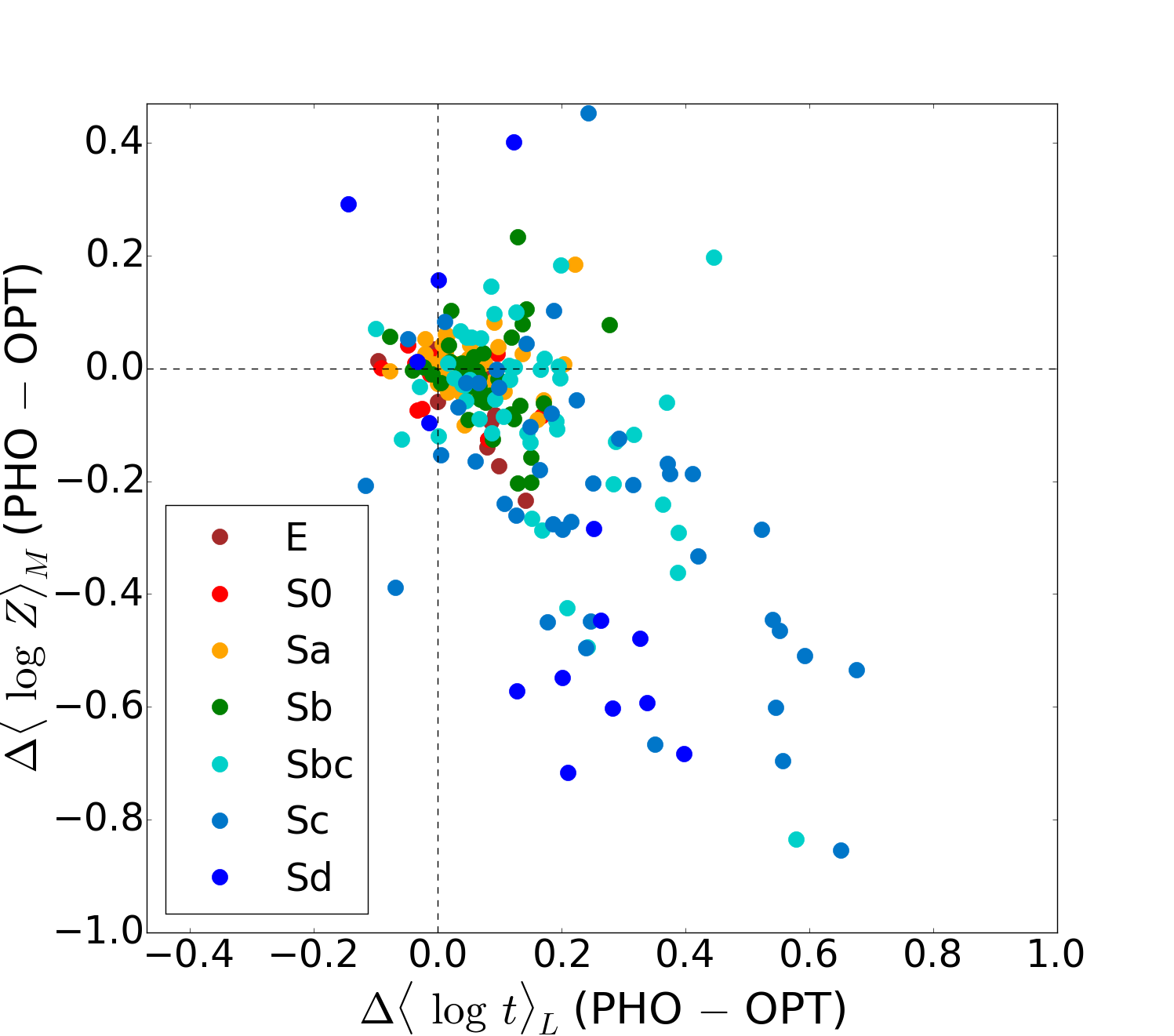}
\caption{Comparison between age and metallicity variations obtained with OPT and PHO fits. 
}
\label{fig:age_z_deg}
\end{figure}
%***FIG***FIG***FIG***FIG***FIG***FIG***FIG***FIG***FIG***FIG***FIG***FIG***FIG***FIG***FIG***FIG

The bottom panels of Fig.\ \ref{fig:CALIFA_prop} compare mean age ($\langle \log t \rangle_L$) and metallicity ($\langle \log Z \rangle_M$) values as estimated from PHO and OPT fits. Here we see an effect which was also detected in the simulations: The difference in $\langle \log t \rangle_L$ is mainly due to the youngest systems, which are somewhat older in PHO than in OPT. By 
virtue of the inter relations between mass, age, metallicity, and morphology, these young galaxies are also late types (Sbc--Sd) of low mass and metallicity.

Metallicities, on the other hand, change in the opposite direction, decreasing from OPT to PHO (bottom right panel of Fig.\ \ref{fig:CALIFA_prop}). Again, the effect is negligible for most galaxies, but can be significant for late type ($M_\star \la 5 \times10^{9} M_\odot$) systems. For these galaxies we obtain $\overline{\Delta} = -0.4$ and $\sigma = 0.26$ dex, whereas for more massive systems the bias and dispersion are just $\overline{\Delta} = -0.04$ and $\sigma = 0.15$ dex.

Fig.\ \ref{fig:age_z_deg} shows that the changes in age and metallicity are anti correlated, reflecting the well known age-metallicity degeneracy (e.g. Worthey 1994). Note, however, that this degeneracy is more frequently studied in the context of early type galaxies and their old stellar populations, while Figs.\ \ref{fig:CALIFA_prop} and \ref{fig:age_z_deg} show that it is late type systems which are affected the most. These are precisely the galaxies for which estimates of the stellar metallicity are harder to obtain, and our results indicate that the UV photometry brings in useful information to improve such estimates.

Finally, we emphasize that even though the anti-correlation stands out in Fig.\ \ref{fig:age_z_deg}, the majority of points cluster around $\Delta \langle \log t \rangle \sim \Delta \langle \log Z \rangle_M \sim 0$, with variations within the uncertainties expected from the simulations. Coupled to the negligible changes in $M_\star$ and $\tau_V$, we conclude that for most galaxies the stellar population properties derived from OPT and PHO fits are consistent with one another. This is an important point to highlight, particularly given that, for obvious reasons, much of this paper is dedicated to mapping the differences between these two approaches.

\subsubsection{Star Formation History}
\label{sec:SFH}

%***FIG***FIG***FIG***FIG***FIG***FIG***FIG***FIG***FIG***FIG***FIG***FIG***FIG***FIG***FIG***FIG
\begin{figure*}
\includegraphics[width=1\textwidth]{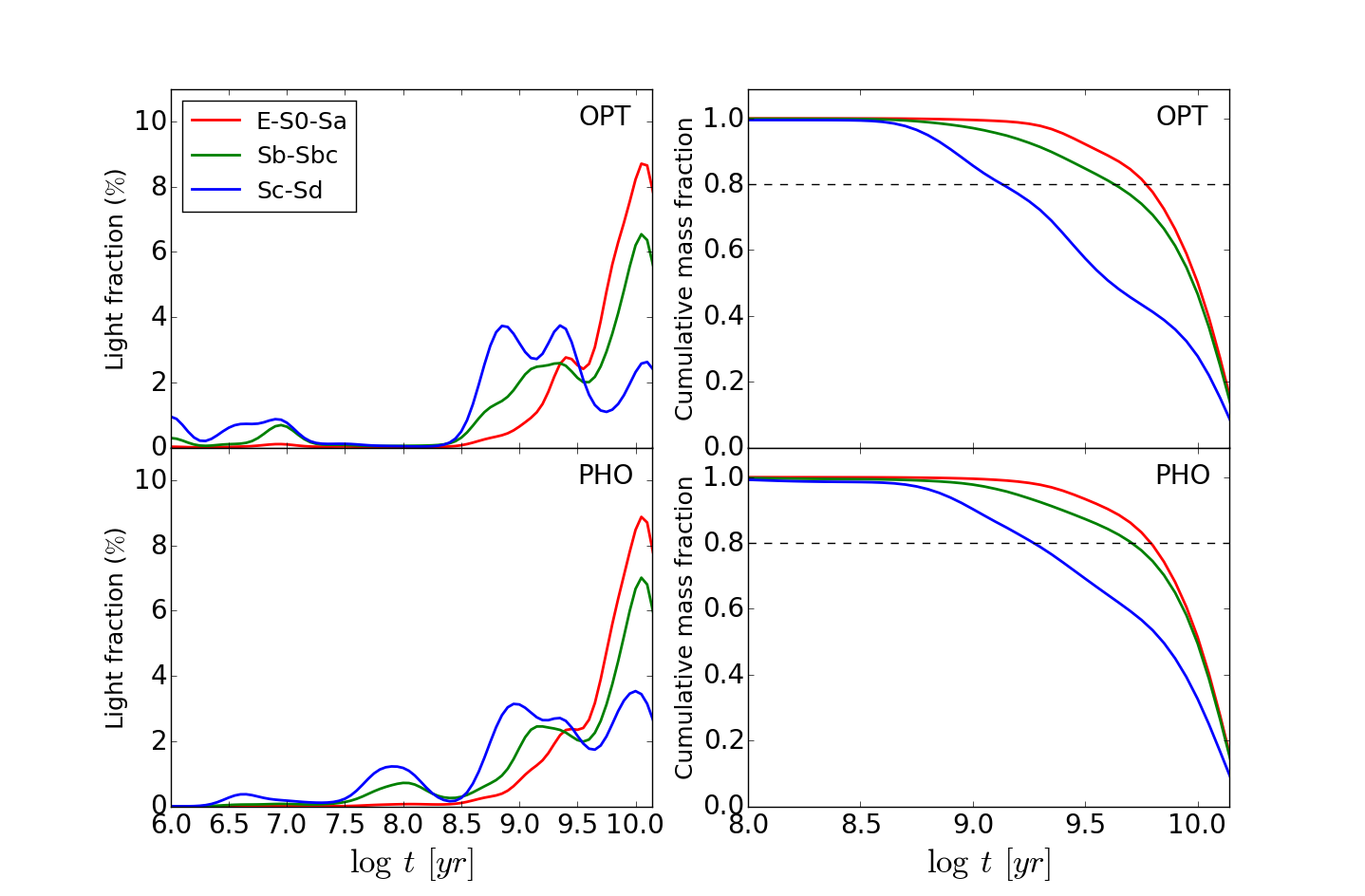}
\caption{
Mean star formation histories of early (red) intermediate (green) and late (blue) type galaxies, as derived from OPT (top panels) and PHO (bottom) fits. Left panels show smoothed versions of the light-fraction population vector ($\vec{x}$), plotted against the lookback time $t$. Right panels show the smoothed cumulative mass  fraction functions, obtained by rescaling the mass converted into stars up to a lookback time $t$ to a 0--1 scale. The 80 per cent line is drawn for reference.
}
\label{fig:SFH_califa}
\end{figure*}
%***FIG***FIG***FIG***FIG***FIG***FIG***FIG***FIG***FIG***FIG***FIG***FIG***FIG***FIG***FIG***FIG

%***FIG***FIG***FIG***FIG***FIG***FIG***FIG***FIG***FIG***FIG***FIG***FIG***FIG***FIG***FIG***FIG
\begin{figure*}
\centering
\includegraphics[width=0.9\textwidth]{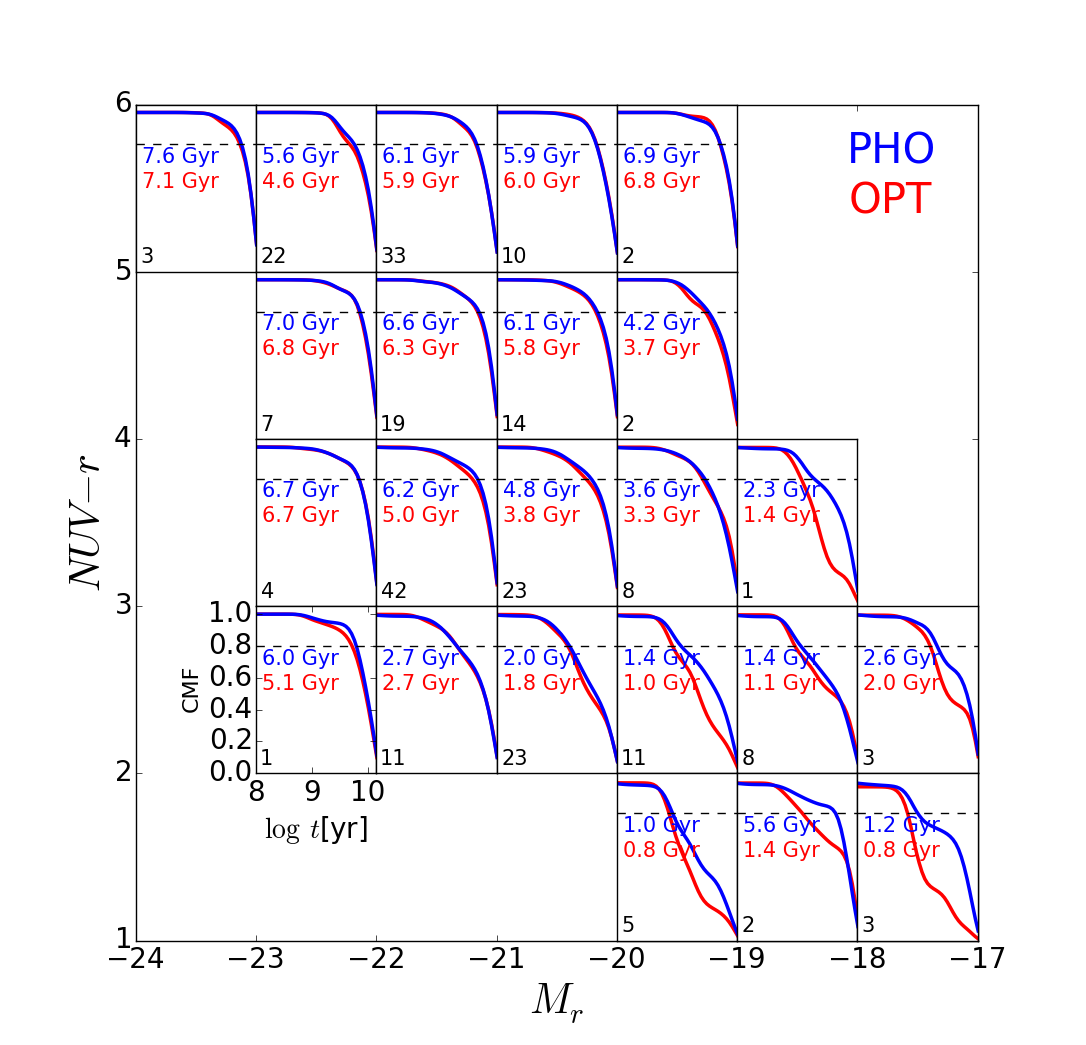}
\caption{Cumulative mass fraction for galaxies in the $NUV - r$ versus $M_{r}$ CMD. Dashed line indicates 80 per cent of the total mass. Blue profiles show the results with PHO fits and red profiles show the results with OPT fits. The number on bottom left of each panel shows the number of galaxies in the bin used to compute the mean curve. The age when 80 per cent of the mass is reached is listed in blue for PHO and red for OPT fits.}
\label{fig:SFHs_on_CMD}
\end{figure*}
%***FIG***FIG***FIG***FIG***FIG***FIG***FIG***FIG***FIG***FIG***FIG***FIG***FIG***FIG***FIG***FIG

The ultimate goal of any stellar population method  is to retrieve the time-dependent SFH. In {\sc starlight} the SFH is encoded in the light ($\vec{x}$) or mass ($\vec{\mu}$) population-vectors. As documented elsewhere (e.g. Cid Fernandes et al.\ 2004, 2014), the individual components of these arrays are highly uncertain, but a coarser description in terms of broad age bins, or, equivalently, smoothed versions of $\vec{x}$ and $\vec{\mu}$, is robust. 

In order to capture general trends in the SFH of our galaxies and how these change from OPT to PHO fits we first smooth the age distribution in $\vec{x}$ and $\vec{\mu}$ by a gaussian in $\log t$ with a  FWHM of 0.7 dex, and average the results in three morphology-defined groups of galaxies: early (E, S0 and Sa), intermediate (Sb and Sbc), and late (Sc, Sd) types. These groups can be seen also as representative of red, green and blue galaxies, respectively. 

The results are shown in Fig.\ \ref{fig:SFH_califa}. Top panels are for OPT fits and bottom ones for PHO. The left plots show the smoothed age distribution in terms of their contribution to the flux at our chosen normalization wavelength $\lambda_0 = 5635$ \AA, while the ones on the right present the cumulative contribution in terms of mass, re-scaling the mass turned into stars onto a 0--1 scale.
Because of the highly non-linear mass-to-light relation of stars, these mathematically  equivalent descriptions of the SFH highlight different aspects of the problem.

In terms of light fractions (left panels in Fig.\ \ref{fig:SFH_califa}), the more relevant differences between OPT and PHO fits are seen for ages $< 300$ Myr. In particular, the $t < 10$ Myr populations seen in OPT fits shift towards $ t \sim 30$--300 Myr when UV constraints are included. This explains the excessive UV flux predicted in OPT fits and the systematically older mean ages obtained with PHO fits. Note, however, that this effect is basically restricted to blue and, to a lesser extent, green galaxies. The SFHs of red (early-type) galaxies do not change significantly from OPT to PHO fits. 
These results clarify the origin of the differences in luminosity weighted mean stellar ages seen in the bottom-left panel of Fig.\ \ref{fig:CALIFA_prop}, where late types stand out as the only ones with significant changes in $\langle \log t \rangle_L$.

In contrast to left panels, the SFHs on the top and bottom right panels of Fig.\ \ref{fig:SFH_califa} are hardly distinguishable. The only visible difference is for blue galaxies, and even then the cumulative mass fractions are very similar. To quantify the differences we compute the age at which the stellar mass has grown to 80 per cent of the total. OPT fits yield $t_{80\%} = 1.3$, 4.2, and 5.9 Gyr for blue, green and red galaxies respectively, while in PHO-fits these values become $t_{80\%} = 1.7$, 5.0, and 5.9 Gyr. This similarity is a consequence of the OPT-PHO changes occurring in relatively young stellar populations, that carry significant light but little mass. 

In Fig.\ \ref{fig:SFHs_on_CMD} we break up the mass assembly histories  in boxes of  $1 \times 1$ mag bins in the $NUV - r$ versus $M_r$ CMD, first studied by Wyder et al.\ (2007). The gradual shift in SFHs towards more recent lookback times as one descends from red to blue bins reflects the strong relation between the $NUV - r$ colour and mean stellar age (further explored in Section \ref{sec:Discussion}),
while the general aging as one moves towards more luminous (smaller $M_r$) bins reflects the well known cosmic downsizing (better appreciated by mentally collapsing the CMD along its $y$-axis). These general tendencies are seen in both OPT and PHO fits. 

Regarding the differences in SFH between OPT and PHO fits, Fig.\ \ref{fig:SFHs_on_CMD} reinforces the conclusion that they are essentially  limited to low mass, blue galaxies, as further confirmed by comparing the values of $t_{80\%}$, listed in blue for PHO and red for OPT fits. A total of 28 of our 260 galaxies reside in CMD bins where $t_{80\%}$ changes by more than 25 per cent (i.e. 0.1 dex). 
Out of our 260 galaxies only 28 reside in CMD bins where $t_{80\%}$ changes by more than 25 per cent (i.e. 0.1 dex). 

%A total of 28 of our 260 galaxies reside in CMD bins where $t_{80\%}$ changes by more than 25 per cent (i.e. 0.1 dex). 

\section{Discussion}
\label{sec:Discussion}

The simulations and empirical results presented above showed that (1) the new code works, i.e. it simultaneously fits an optical spectrum and UV photometry, as designed to, and (2) OPT and PHO fits only differ relevantly for low-mass, late type-galaxies, whose mean ages become somewhat older while their mean metallicities tend to decrease. 

In this section we discuss how the addition of UV constraints affects previously known results. We first examine empirical relations between mean stellar age and observables such as colours and the 4000 \AA\ break (Section \ref{sec:AC_col}). The scatter in these relations provides an indirect way of assessing whether PHO fits are more reliable than OPT ones, as intuitively expected. We then revisit the relation between stellar metallicity and (a) mass, and (b) mass surface  density, comparing our own previous OPT-based CALIFA results with those obtained with our CALIFA + GALEX sample.

\subsection{Empirical age indicators}
\label{sec:AC_col}

%***FIG***FIG***FIG***FIG***FIG***FIG***FIG***FIG***FIG***FIG***FIG***FIG***FIG***FIG***FIG***FIG
\begin{figure*}
\centering
\includegraphics[width=0.85\textwidth]{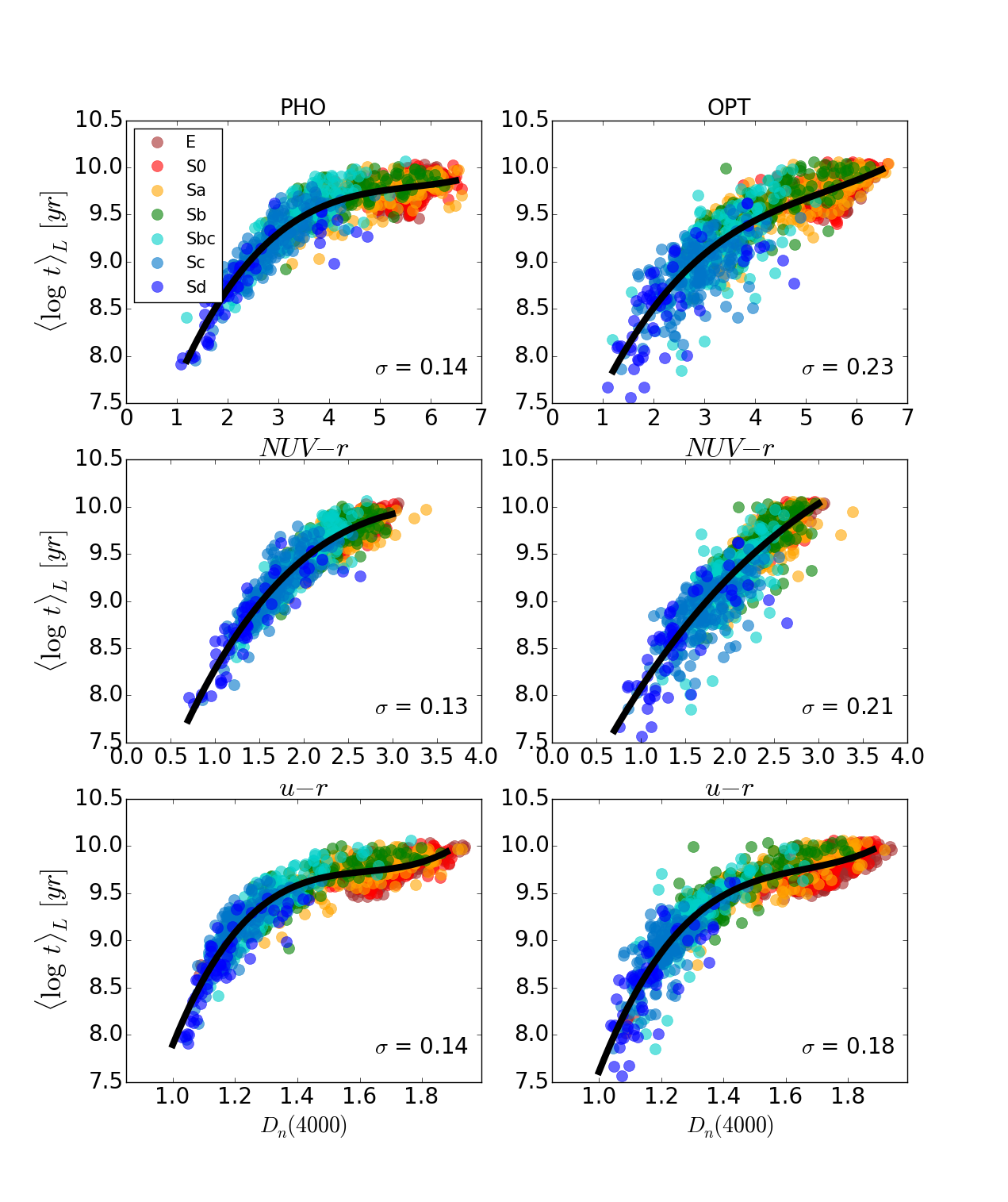}
\caption{
Empirical correlations between the {\sc starlight}-derived luminosity weighted mean stellar age and the observed $NUV - r$ (top panels), $u - r$ (middle), and $D_{n}(4000)$ (bottom). Left panels are for PHO-fits and right ones for OPT fits. The black lines show polynomial fits to the points (see Table \ref{table:pol_fit}), and $\sigma$ denotes the rms dispersion with respect to these lines. Points are coloured by morphological type following the palette in Fig.\ \ref{fig:CMD_NUV_r}. The points in these diagrams come from four different radial extractions for each of our 260 galaxies.
}
\label{fig:NUV-r_age_Morph}
\end{figure*}
%***FIG***FIG***FIG***FIG***FIG***FIG***FIG***FIG***FIG***FIG***FIG***FIG***FIG***FIG***FIG***FIG

Because PHO fits incorporate more constraints one tends to regard their output results as superior to those derived from OPT fits. Still, it would be nice to have some independent way of evaluating which approach produces better results.

We perform this judgement by comparing the scatter in empirical relations such as those shown in Fig.\ \ref{fig:NUV-r_age_Morph}, where we relate our {\sc starlight}-derived luminosity weighted mean (log) stellar age to observable properties.
Top panels show the correlation between $\langle \log t \rangle_L$ and the $NUV - r$ colour, while in the middle and bottom panels the x-axis is replaced by $u-r$ and the 4000 \AA\ break index (Balogh et al.\ 1999), both purely optical properties. 
Left and right columns correspond to PHO and OPT fits, respectively. The data used in this figure come from the extended data set discussed in Section \ref{sec:CALIFAGALEXsample}, containing both integrated properties and values derived from four different spatial extractions. The improved statistics of this larger sample serves to better delineate the correlations.

A simple visual inspection suffices to conclude that the OPT-based relations are more dispersed than those based on PHO fits. This is not really unexpected in the case of the top panels, since OPT fits are completely oblivious of the $NUV - r$ colour, whereas PHO fits do take this information into account. In the middle and bottom panels, however, the x-axis represents properties which are known to both OPT and PHO fits. Still, the relations between $\langle \log t \rangle_L$ and $u-r$ and between $\langle \log t \rangle_L$ and $D_n(4000)$ are visibly better defined with PHO mean ages than with OPT ones. This improvement can be quantified by comparing the $\sigma$ values given in each panel  of Fig.\ \ref{fig:NUV-r_age_Morph}, which represent the dispersion around the polynomial fits shown as solid lines and whose coefficients are given in Table \ref{table:pol_fit}. The scatter in OPT-based relations is almost twice as large as for PHO-based ones.
As is evident from the cyan-blue color of most outliers, this reduced scatter occurs because of late-type galaxies.

In short, besides taking more observational constraints into consideration, PHO fits produce better behaved mean stellar ages than OPT fits, in the sense that they correlate better (less scatter) with classical observable age indicators. PHO thus outplay OPT in this qualitative assessment.

\subsection{Less age-metallicity-extinction degeneracies with UV data}
\label{sec:Age-Z-deg}

%***FIG***FIG***FIG***FIG***FIG***FIG***FIG***FIG***FIG***FIG***FIG***FIG***FIG***FIG***FIG***FIG
\begin{figure}
\includegraphics[width=0.52\textwidth]{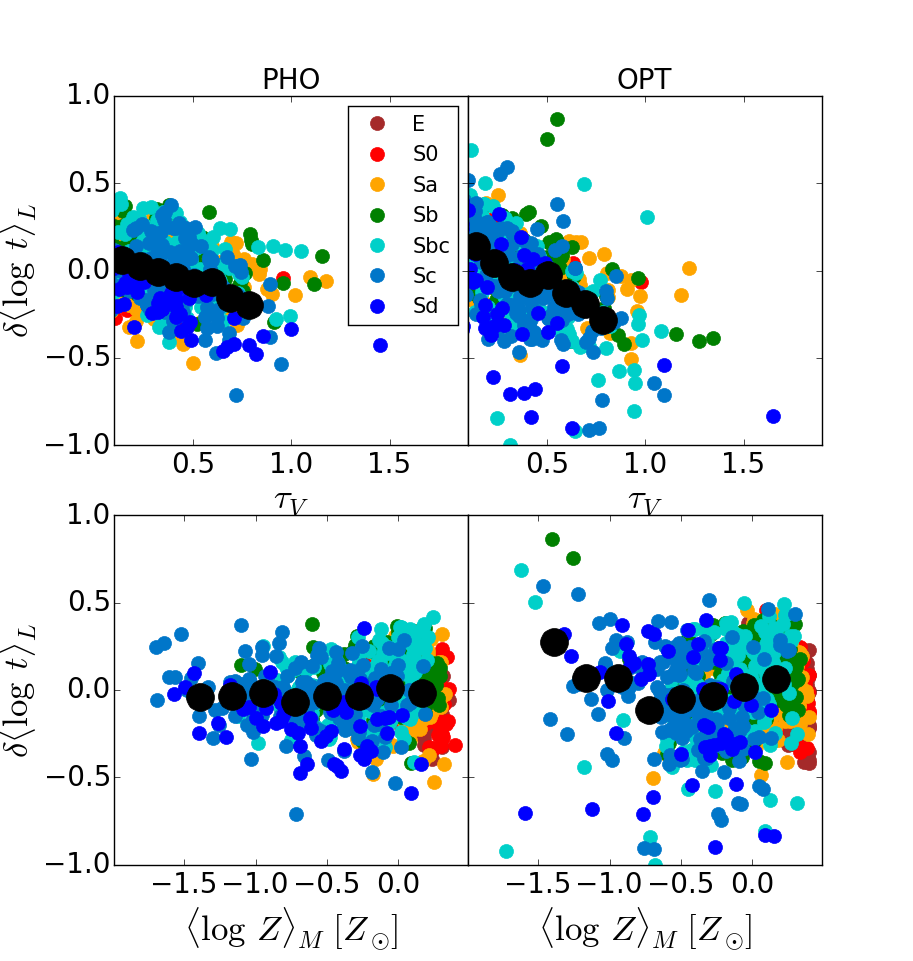}
\caption{Residual from the $\langle\log\, t\rangle_L(NUV-r)$ relation (solid line in the top panels of Fig.\ \ref{fig:NUV-r_age_Morph})
plotted against dust optical depth (top) and mean stellar metallicity (bottom), for both PHO (left) and OPT (right) fits.
Black points trace the mean values for bins in the $x$-axis.
}
\label{fig:NUV-r_disp}
\end{figure}
%***FIG***FIG***FIG***FIG***FIG***FIG***FIG***FIG***FIG***FIG***FIG***FIG***FIG***FIG***FIG***FIG

Age and metallicity are more sensitive to UV than optical colours (Yi et al.\ 2004), and previous studies suggest that combining optical and UV provides better estimates of ages and can effectively break (or at least mitigate the effects of) the age-metallicity degeneracy (Yi 2003; Kaviraj et al.\ 2007). Indeed, we have just seen that a combined optical+UV analysis produces more accurate mean age estimates than a purely optical one for Sbc--Sd galaxies.

Because of the known covariances amongst stellar population properties, more accurate ages should also lead to more accurate metallicity and reddening estimates. Without independent fiducial values to compare to it is not possible to directly verify if this is indeed the case. With this caveat in mind, we try to shed some light onto this issue by examining the origin of the dispersion in the empirical relations shown in  Fig.\ \ref{fig:NUV-r_age_Morph}.

%***TAB***TAB***TAB***TAB***TAB***TAB***TAB***TAB***TAB***TAB***TAB***TAB***TAB***TAB***TAB
\begin{table*}
\centering % used for centering table
\begin{tabular}{c r r r r r} % centered columns (4 columns)
\multicolumn{3}{|c|}{Empirical age indicators: Polynomial fits} \\
\hline %inserts double horizontal lines
Age indicator ($x$) & \multicolumn{1}{|c|}{PHO} & \multicolumn{1}{|c|}{OPT}\\ %[0.5ex] % inserts table
%heading
\hline % inserts single horizontal line
%r & 0.000 \pm 0.008 & 0.000 \pm 0.004 & 0.000 \pm 0.002 & 0.000 \pm 0.000 & 
%0.000 \pm 0.000 \\
$NUV - r$  &  $0.0176x^{3} - 0.2991x^{2} + 1.7402x + 6.3182$  & 
$0.0150x^{3} - 0.2421x^{2} + 1.4965x + 6.3654$\\
$u - r$  & $0.0544x^{3} - 0.6761x^{2} + 2.8225x + 6.0703$ &
$ 0.0242x^{3} - 0.3487x^{2} + 2.0571x + 6.3474$ \\
%$\chi^2_ {\rm SPEC}$  & $1360.50 \pm 56.57 $ & $ 1361.07 \pm 55.07 $ & $ 1355.96 \pm %56.61 $&  $1355.42 \pm 54.01 $&   $1361.36 \pm 58.47$\\
$D_{n}(4000)$ & $6.86773x^{3} - 33.3609x^{2} + 54.3533x - 19.9681$ & 
$5.9043x^{3} - 29.4365x^{2} + 49.5827x - 18.4440$ \\
%adev ($\%$) & $17.38 \pm 0.46 $& $8.09 \pm 0.18 $& $3.98 \pm 0.09$ & 
%$1.59 \pm 0.03$ & $0.79 \pm 0.02 $\\
\hline %inserts single line
\end{tabular}
\centering
\caption{Polynomial fits for empirical relations between mean (luminosity weighted) stellar age obtained with {\sc starlight} and different observables (see Fig.\ \ref{fig:NUV-r_age_Morph}): $\langle \log t / {\rm yr} \rangle_L = a x^3 + b x^2 + c x + d$, where $x = NUV - r$, $u-r$, or $D_n(4000)$. 
}
\label{table:pol_fit} 
\end{table*}
%***TAB***TAB***TAB***TAB***TAB***TAB***TAB***TAB***TAB***TAB***TAB***TAB***TAB***TAB***TAB

In Fig. \ref{fig:NUV-r_disp} we use the  best fit $\langle\log t\rangle_L(NUV - r)$ relation given in  Table \ref{table:pol_fit} to investigate what third variable is responsible for the scatter in mean stellar age at fixed UV-optical colour. The figure shows the $\delta  \langle\log t\rangle_L  \equiv \langle\log t\rangle_L  - \langle\log t\rangle_L({NUV-r})$ residual as a function of $\tau_V$ (top panels) and $\langle \log Z \rangle_M$ (bottom), for both PHO (left) and OPT fits (right). A strong anti correlation with $\tau_V$ is seen for both PHO and OPT fits, with $\delta  \langle\log t\rangle_L$ becoming increasingly negative for increasing $\tau_V$.  This anti correlation is expected given the long wavelength baseline in the  $NUV - r$ colour, which makes it susceptible to $\tau_V$. The relation is steeper and visibly more dispersed for OPT fits.

For OPT fits, metallicity also seems to play a role in the dispersion around the  $\langle \log t\rangle_L(NUV-r)$ relation, as inferred from the bottom right panel of Fig.\ \ref{fig:NUV-r_disp}. Besides a significant dispersion, the trend of increasing $\delta  \langle\log t\rangle_L$ for decreasing $\langle \log Z \rangle_M$ at low metallicities is qualitatively consistent with what one expects from the age-metallicity degeneracy.
 PHO fits, on the other hand, produce a $\sim$ flat relation between $\delta  \langle\log t\rangle_L$ and $\langle \log Z \rangle_M$ (bottom left panel), indicating that the inclusion of UV data indeed minimizes the effect of the age-metallicity degeneracy with respect to fits considering only the optical spectrum.
This qualitative assessment therefore reinforces our conclusion that PHO fits produce better constrained physical properties.

\subsection{The stellar mass-metallicity relation}

%***FIG***FIG***FIG***FIG***FIG***FIG***FIG***FIG***FIG***FIG***FIG***FIG***FIG***FIG***FIG***FIG
\begin{figure}
\centering
\includegraphics[width=0.5\textwidth]{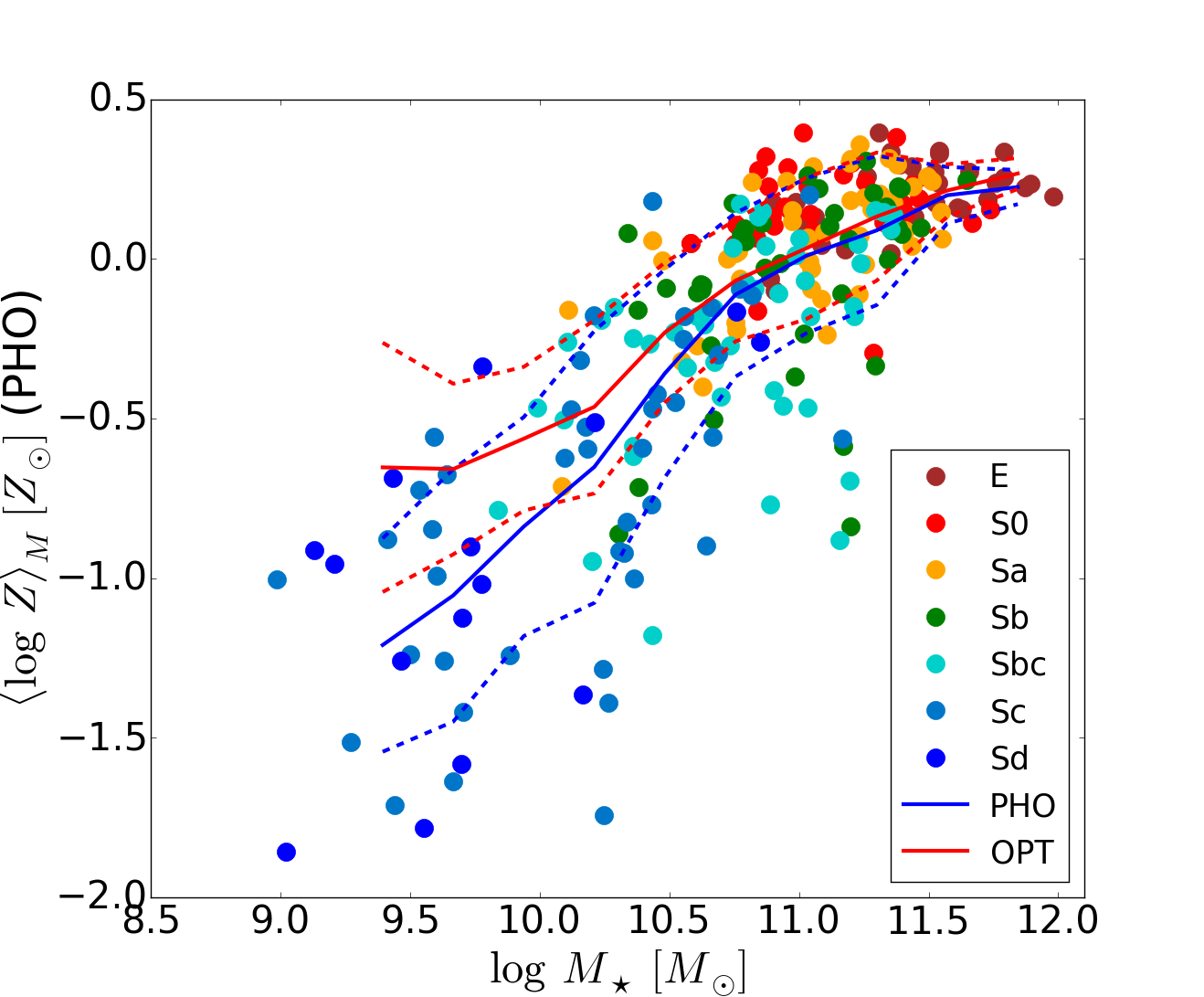}
\caption{The mass-metallicity relation obtained with PHO fits. Blue lines indicate the mean profile with $\pm \sigma$ standard deviation (solid and dashed lines, respectively). Red lines show the mean mass-metallicity relation profile obtained with OPT fits with $\pm \sigma$ standard deviation (solid and dashed lines, respectively).
}
\label{fig:M-Z-relation}
\end{figure}
%***FIG***FIG***FIG***FIG***FIG***FIG***FIG***FIG***FIG***FIG***FIG***FIG***FIG***FIG***FIG***FIG

%***FIG***FIG***FIG***FIG***FIG***FIG***FIG***FIG***FIG***FIG***FIG***FIG***FIG***FIG***FIG***FIG
\begin{figure}
\centering
\includegraphics[width=0.5\textwidth]{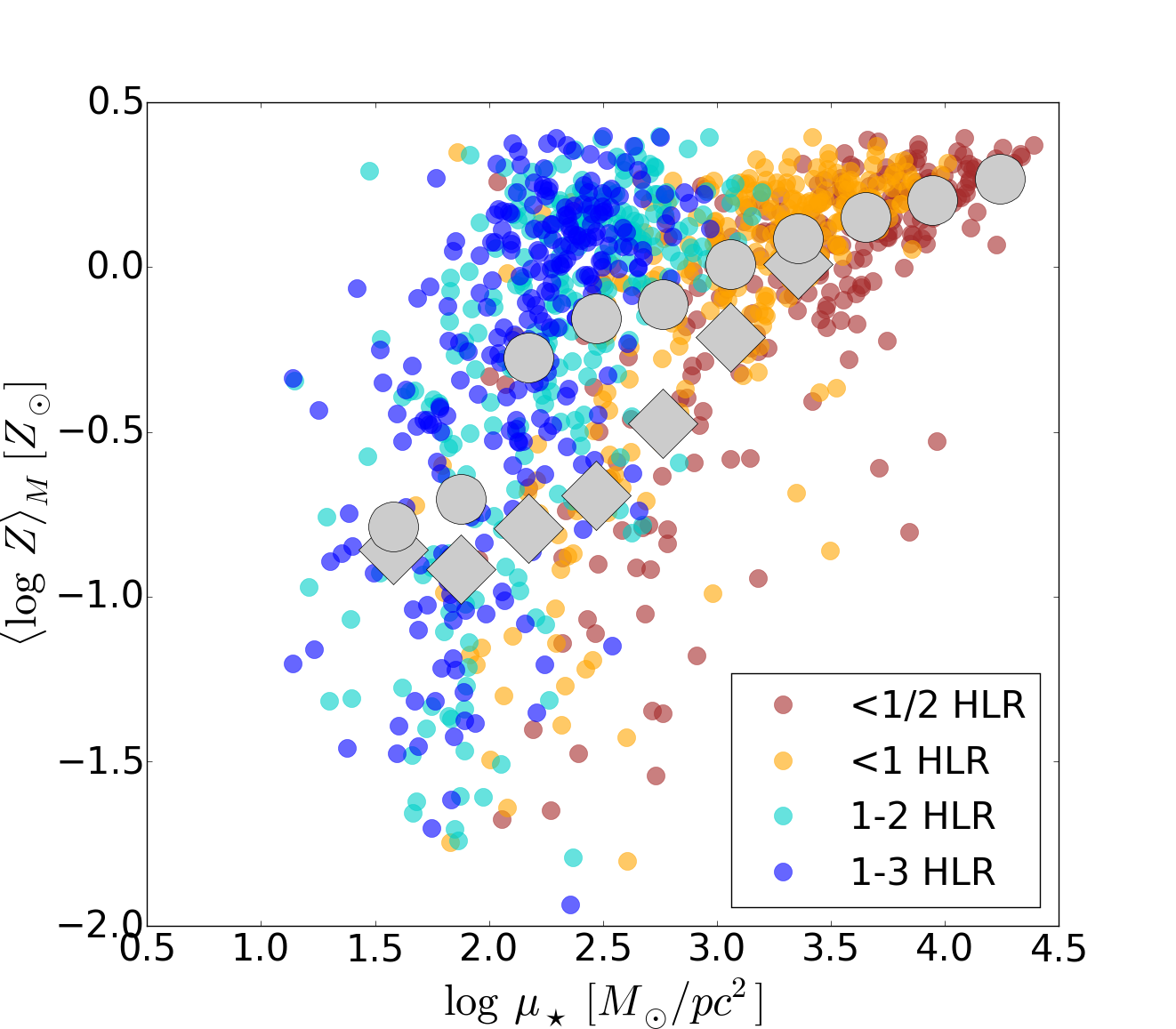}
\caption{
Local stellar metallicity versus the local stellar mass surface density obtained with PHO fits. Colours code results obtained for different radial extractions (in units of the optical Half Light Radius). Grey circles track the $\mu_{\star}$-binned average $Z_{\star}(\mu_{\star})$ relation. Grey diamonds track the relation for Sc and Sd galaxies. 
}
\label{fig:muZ-relation}
\end{figure}
%***FIG***FIG***FIG***FIG***FIG***FIG***FIG***FIG***FIG***FIG***FIG***FIG***FIG***FIG***FIG***FIG

One of the most important correlations in galaxy evolution work is the one between mass and metallicity (Tremonti et al.\ 2004; Gallazzi et al.\ 2005). In GD14 we have presented a {\sc starlight}-based study on the relations between stellar metallicity and mass in CALIFA galaxies, both on global (i.e. galaxy-wide) and local (spatially resolved) scales. In this section we examine whether and how the mass-metallicity (MZR) and surface mass density-metallicity ($\mu$ZR) relations change with the addition of UV data to spectroscopic data.\footnote{Because of the differences in the SSP models used in the synthesis, and, less importantly, in the sample, our results should not be directly compared to those of GD14. The comparison between PHO and OPT fits presented here, however, is based on equal ingredients and therefore is meaningful.} 

Our comparative analysis of PHO and OPT fits showed that changes in the mean stellar metallicity are only relevant and systematic for late type galaxies (Fig.\ \ref{fig:CALIFA_prop}), whose $\langle \log Z \rangle_M$ values decrease by $\sim 0.4$ dex on average. These presumably more accurate estimates are particularly welcome for these low-mass, star-forming galaxies, where the inherent difficulties in estimating stellar metallicities are aggravated by the almost featureless continuum of their hot stars,  which heavily dilute the absorption lines carrying information on $Z$. 

We can thus anticipate that changes in the MZR should be restricted to the low $Z$, low $M_\star$ end of the relation. This is confirmed in  Fig.\ \ref{fig:M-Z-relation}, where we show the PHO-based MZR for the 260 galaxies in our combined CALIFA+GALEX sample. The blue lines show the smoothed mean relation (solid line) and the corresponding $\pm 1 \sigma$ range (dashed). The OPT-based MZR for this sample is shown by the red line. As expected, the two are essentially identical at the high mass end, diverging towards low masses, with PHO fits reaching lower metallicities.

Fig.\ \ref{fig:muZ-relation}  presents the $\mu$ZR as derived from our optical+UV {\sc starlight} analysis. Small points are coloured according to the four spatial extractions discussed in Section \ref{sec:CALIFAGALEXsample}. Despite the much coarser spatial resolution, the same dichotomy identified by GD14 is seen in our PHO-version of the local $\mu$ZR, with inner regions (brown and orange points) exhibiting a visibly flatter $\mu$ZR than outer ones (cyan and blue). In fact, because of the stretched  $\langle \log Z \rangle_M$  scale, this dichotomy becomes ever clearer than in a purely optical analysis. Reinforcing the results of GD14, we find that the mean stellar metallicity is strongly related to the local density within galactic discs, while in spheroids $\mu_\star$ does not seem to play a major role in controlling chemical evolution. 

The large grey circles in Fig.\ \ref{fig:muZ-relation} show a smoothed mean $\mu$ZR, obtained by averaging $\langle \log Z \rangle_M$-values in bins of $\mu_\star$. This mean $\mu$ZR reflects the dual behaviour of the individual points, with an initially steep relation becoming weaker (flatter) above $\sim 10^{2.5} M_\odot \, \text{pc}^{-2}$. Diamonds in Fig.\ \ref{fig:muZ-relation} indicate the mean $\mu$ZR obtained when restricting the sample to Sc and Sd galaxies. The metal poor, dense, inner (red and orange) zones of these late type galaxies account for nearly all of the outliers at $\mu_\star \ga 10^3 M_\odot \, \text{pc}^{-2}$. As suggested by GD14, this may be related to the $\sim$ bulgelessness of these systems making their inner regions behave in a disc-like way in the stellar metallicity versus surface mass density space.

\section{Summary}
\label{sec:Summary}

As part of an effort to extend the capabilities of the {\sc starlight} code, an extensively used tool to estimate stellar population properties of galaxies out of detailed $\lambda$-by-$\lambda$ spectral fits, a new version was developed with the specific purpose of including photometric data in the fit, thereby incorporating more observational constraints in the analysis.

This paper was dedicated to the presentation and validation of this extended {\sc starlight} by means of simulations and data comprising a 3700--7000 \AA\ optical spectrum plus NUV and FUV magnitudes. The observational data consist of  spatially matched CALIFA datacubes and GALEX images for a varied sample of 260 galaxies of all types from E to Sd. Both simulated and actual data were fit using (a) only the optical spectrum (OPT), and (b) the full optical+UV data (PHO), thus allowing us to gauge the practical benefits of adding UV constraints to an optical full spectral fitting analysis.

Our main results can be summarized as follows:

\begin{enumerate}

\item By themselves, optical spectral fits are poor predictors of the UV properties, with errors of the order of 0.5 mag and a tendency to overpredict the fluxes, even for high signal-to-noise. This happens because optically insignificant young stellar populations can dominate the emission in the UV, so that even minor errors in the estimation of their optical contribution translate into large errors in the UV. 

\item Besides matching the input UV data to within the errors, the new optical+UV fits reduce the uncertainties in the derived stellar properties. 

\item Applying the code to a combination of CALIFA+GALEX data we find that including UV photometry in the fits better constrains the contribution of stellar populations younger than  $\sim 300$ Myr old. PHO fits tend to replace  $\la 30$ Myr components by populations in the neighbourhood of 100 Myr (i.e. from O and B stars to B and A).

\item Despite their poor performance in predicting the UV fluxes, for nearly 90\% of our sample OPT fits yield stellar population properties which agree with those obtained with PHO fits to within the expected uncertainties. The differences are exclusively found in low-mass, late-type galaxies, precisely the systems where, because of their significant $\la 300$ Myr population, one would expect the addition of UV constraints to play a more relevant role. 

\item For Sc and Sd galaxies with $M_{\star} < 5 \times 10^{9} M_{\odot}$ we find that an optical+UV analysis yields older ages and lower metallicities than those derived  with purely optical fits. These changes imply a steepening of the relations between stellar metallicity and mass (MZR) and surface density ($\mu$ZR) at the low $Z$ end, making the dual disc and spheroid behaviours even clearer than previously reported with OPT-based studies by our own group.

\item Empirical relations between our (luminosity weighted) mean stellar age and observables such as the 4000 \AA\ break, and UV and optical colours, are all less dispersed for PHO than for OPT fits, which indicates that the inclusion of UV constraints helps mitigating degeneracies between age, dust and metallicity.

\end{enumerate}

All the experiments conducted in this paper were based on optical spectra + UV photometry, so future tests of the new code should explore other combinations, like optical spectra + near IR photometry, and others. In any case, at least in the regime tested here, the extended {\sc starlight} is ready to extract stellar population properties out of combinations of photometry and spectroscopy. SDSS spectra and GALEX photometry comprise a natural dataset to explore, although it demands care in dealing with aperture mismatches, an issue which we   circumvented in this study through the use of integral field data. In parallel to these studies, and despite the progress already achieved with the incorporation of photometric constraints, further upgrades to {\sc starlight} are desirable, and implementing a more realistic modelling of dust effects is top on the priority list. Differential extinction, in particular, may well have a significant impact in studies of star-forming systems,
%(including some of the astrophysical results reported in this very study),
 and work is underway to tackle this issue.

\section*{ACKNOWLEDGMENTS}

CALIFA is the first legacy survey carried out at Calar Alto, and we  thank the IAA-CSIC and MPI-MPG as major partners of the observatory, and CAHA itself, for the unique access to telescope time and support in manpower and infrastructures. We also thank the CAHA staff for the dedication to this project. Support from the Spanish Ministerio de Econom\'ia y Competitividad, through projects AYA2014-57490-P, AYA2010-15081 (PI RGD), Junta de Andaluc\'ia FQ1580 (PI RGD), AYA2010-22111-C03-03, AYA2010-10904E (SFS) and short-term research FPI program grants EEBB-I-2013-07071 and EEBB-I-2014-08601. SFS thanks the CONACYT-125180 and DGAPA-IA100815 projects for providing him support in this
study. We also thank the Viabilidad, Dise\~no, Acceso y Mejora funding program, ICTS-2009-10, for funding the data acquisition of this project. Support from the Brazilian Science without borders program, as well as CNPq and CAPES is duly acknowledged. This research made use of Montage. It is funded by the National Science Foundation under Grant Number ACI-1440620, and was previously funded by the National Aeronautics and Space Administration's Earth Science Technology Office, Computation Technologies Project, under Cooperative Agreement Number NCC5-626 between NASA and the California Institute of Technology.


\begin{thebibliography}{99}
\bibitem[\protect\citeauthoryear{Balogh et al.}{1999}]{1999ApJ...527...54B} 
Balogh M.~L., Morris S.~L., Yee H.~K.~C., Carlberg R.~G., Ellingson E., 
1999, ApJ, 527, 54 

\bibitem[\protect\citeauthoryear{Barway et al.}{2013}]{2013MNRAS.432..430B} 
Barway S., Wadadekar Y., Vaghmare K., Kembhavi A.~K., 2013, MNRAS, 432, 430

\bibitem[\protect\citeauthoryear{Bruzual 
\& Charlot}{2003}]{2003MNRAS.344.1000B} Bruzual G., Charlot S., 2003, MNRAS, 344, 1000 

\bibitem[\protect\citeauthoryear{Calzetti, Kinney, 
\& Storchi-Bergmann}{1994}]{1994ApJ...429..582C} Calzetti D., Kinney A.~L., Storchi-Bergmann T., 1994, ApJ, 429, 582

\bibitem[\protect\citeauthoryear{Calzetti et 
al.}{2000}]{2000ApJ...533..682C} Calzetti D., Armus L., Bohlin R.~C., 
Kinney A.~L., Koornneef J., Storchi-Bergmann T., 2000, ApJ, 533, 682

\bibitem[\protect\citeauthoryear{Charlot 
\& Fall}{2000}]{2000ApJ...539..718C} Charlot S., Fall S.~M., 2000, ApJ, 539, 718

\bibitem[\protect\citeauthoryear{Cid Fernandes et 
al.}{2004}]{2004MNRAS.355..273C} Cid Fernandes R., Gu Q., Melnick J., 
Terlevich E., Terlevich R., Kunth D., Rodrigues Lacerda R., Joguet B., 
2004, MNRAS, 355, 273

\bibitem[\protect\citeauthoryear{Cid Fernandes et 
al.}{2005}]{2005MNRAS.358..363C} Cid Fernandes R., Mateus A., Sodr{\'e} L., 
Stasi{\'n}ska G., Gomes J.~M., 2005, MNRAS, 358, 363

\bibitem[\protect\citeauthoryear{Cid Fernandes}{2006}]{2006BAAA...49..228C} 
Cid Fernandes R., 2006, BAAA, 49, 228

\bibitem[\protect\citeauthoryear{Cid Fernandes 
\& Gonz{\'a}lez Delgado}{2010}]{2010MNRAS.403..780C} Cid Fernandes R., Gonz{\'a}lez Delgado R.~M., 2010, MNRAS, 403, 780

\bibitem[\protect\citeauthoryear{Cid Fernandes et 
al.}{2011}]{2011MNRAS.413.1687C} Cid Fernandes R., Stasi{\'n}ska G., Mateus 
A., Vale Asari N., 2011, MNRAS, 413, 1687

\bibitem[\protect\citeauthoryear{Cid Fernandes et 
al.}{2013}]{2013A&A...557A..86C} Cid Fernandes R., et al., 2013, A\&A, 557, A86

\bibitem[\protect\citeauthoryear{Cid Fernandes et 
al.}{2014}]{2014A&A...561A.130C} Cid Fernandes R., et al., 2014, A\&A, 561, A130

\bibitem[\protect\citeauthoryear{Gallazzi et 
al.}{2005}]{2005MNRAS.362...41G} Gallazzi A., Charlot S., Brinchmann J., 
White S.~D.~M., Tremonti C.~A., 2005, MNRAS, 362, 41 

\bibitem[\protect\citeauthoryear{Garc{\'{\i}}a-Benito et 
al.}{2015}]{2015A&A...576A.135G} Garc{\'{\i}}a-Benito R., et al., 2015, A\&A, 576, A135

\bibitem[\protect\citeauthoryear{Gonz{\'a}lez Delgado et 
al.}{2004}]{2004ApJ...605..127G} Gonz{\'a}lez Delgado R.~M., Cid Fernandes 
R., P{\'e}rez E., Martins L.~P., Storchi-Bergmann T., Schmitt H., Heckman 
T., Leitherer C., 2004, ApJ, 605, 127 

%\bibitem[\protect\citeauthoryear{Gonz{\'a}lez Delgado et al.}{2014}]{2014A&A...562A..47G} Gonz{\'a}lez Delgado R.~M., et al., 2014, A\&A, 562, A47

\bibitem[\protect\citeauthoryear{Gonz{\'a}lez Delgado et 
al.}{2014}]{2014ApJ...791L..16G} Gonz{\'a}lez Delgado R.~M., et al., 2014, 
ApJ, 791, L16

\bibitem[\protect\citeauthoryear{Gonz{\'a}lez Delgado et 
al.}{2015}]{2015A&A...581A.103G} Gonz{\'a}lez Delgado R.~M., et al., 2015, A\&A, 581, A103

\bibitem[\protect\citeauthoryear{Husemann et 
al.}{2013}]{2013A&A...549A..87H} Husemann B., et al., 2013, A\&A, 549, A87

\bibitem[\protect\citeauthoryear{Kaviraj et 
al.}{2007}]{2007ApJS..173..619K} Kaviraj S., et al., 2007, ApJS, 173, 619

\bibitem[\protect\citeauthoryear{Kaviraj et 
al.}{2007}]{2007MNRAS.381L..74K} Kaviraj S., Rey S.-C., Rich R.~M., Yoon 
S.-J., Yi S.~K., 2007, MNRAS, 381, L74

\bibitem[\protect\citeauthoryear{Kennicutt}{1998}]{1998ARA&A..36..189K} Kennicutt R.~C., Jr., 1998, ARA\&A, 36, 189 

\bibitem[\protect\citeauthoryear{Martin et al.}{2005}]{2005ApJ...619L...1M} 
Martin D.~C., et al., 2005, ApJ, 619, L1

\bibitem[\protect\citeauthoryear{Noll et 
al.}{2009}]{2009A&A...507.1793N} Noll S., Burgarella D., Giovannoli E., Buat V., Marcillac D., Mu{\~n}oz-Mateos J.~C., 2009, A\&A, 507, 1793 

\bibitem[\protect\citeauthoryear{P{\'e}rez et 
al.}{2013}]{2013ApJ...764L...1P} P{\'e}rez E., et al., 2013, ApJ, 764, L1

\bibitem[\protect\citeauthoryear{Roth et al.}{2005}]{2005PASP..117..620R} 
Roth M.~M., et al., 2005, PASP, 117, 620 

\bibitem[\protect\citeauthoryear{Salim et al.}{2007}]{2007ApJS..173..267S} 
Salim S., et al., 2007, ApJS, 173, 267

\bibitem[\protect\citeauthoryear{S{\'a}nchez et 
al.}{2012}]{2012A&A...538A...8S} S{\'a}nchez S.~F., et al., 2012, A\&A, 538, A8

\bibitem[\protect\citeauthoryear{Schawinski et 
al.}{2007}]{2007ApJS..173..512S} Schawinski K., et al., 2007, ApJS, 173, 
512 

\bibitem[\protect\citeauthoryear{Schawinski et 
al.}{2014}]{2014MNRAS.440..889S} Schawinski K., et al., 2014, MNRAS, 440, 
889

\bibitem[\protect\citeauthoryear{Schiminovich et 
al.}{2007}]{2007ApJS..173..315S} Schiminovich D., et al., 2007, ApJS, 173, 
315

\bibitem[\protect\citeauthoryear{Taylor et al.}{2011}]{2011MNRAS.418.1587T} 
Taylor E.~N., et al., 2011, MNRAS, 418, 1587  

\bibitem[\protect\citeauthoryear{Tremonti et 
al.}{2004}]{2004ApJ...613..898T} Tremonti C.~A., et al., 2004, ApJ, 613, 
898

\bibitem[\protect\citeauthoryear{Verheijen et 
al.}{2004}]{2004AN....325..151V} Verheijen M.~A.~W., Bershady M.~A., 
Andersen D.~R., Swaters R.~A., Westfall K., Kelz A., Roth M.~M., 2004, AN, 
325, 151

\bibitem[\protect\citeauthoryear{Walcher et 
al.}{2011}]{2011Ap&SS.331....1W} Walcher J., Groves B., Budav{\'a}ri T., Dale D., 2011, Ap\&SS, 331, 1

\bibitem[\protect\citeauthoryear{Walcher et 
al.}{2014}]{2014A&A...569A...1W} Walcher C.~J., et al., 2014, A\&A, 569, A1

\bibitem[\protect\citeauthoryear{Worthey}{1994}]{1994ApJS...95..107W} 
Worthey G., 1994, ApJS, 95, 107

\bibitem[\protect\citeauthoryear{Wyder et al.}{2007}]{2007ApJS..173..293W} 
Wyder T.~K., et al., 2007, ApJS, 173, 293

\bibitem[\protect\citeauthoryear{Yi}{2003}]{2003ApJ...582..202Y} Yi S.~K., 
2003, ApJ, 582, 202

\bibitem[\protect\citeauthoryear{Yi et al.}{2004}]{2004MNRAS.349.1493Y} Yi 
S.~K., Peng E., Ford H., Kaviraj S., Yoon S.-J., 2004, MNRAS, 349, 1493

\end{thebibliography}
\end{document}